\documentclass{jpp}
\usepackage{graphicx}
\usepackage{natbib}
\usepackage[T1]{fontenc}
\usepackage{bm}
\usepackage{color}
\usepackage{amsmath}
\usepackage{amssymb}
\usepackage{amsfonts}
\usepackage{courier}
\usepackage{txfonts}
\ifCUPmtlplainloaded \else
  \checkfont{eurm10}
  \iffontfound
    \IfFileExists{upmath.sty}
      {\typeout{^^JFound AMS Euler Roman fonts on the system,
                   using the 'upmath' package.^^J}%
       \usepackage{upmath}}
      {\typeout{^^JFound AMS Euler Roman fonts on the system, but you
                   dont seem to have the}%
       \typeout{'upmath' package installed. JPP.cls can take advantage
                 of these fonts, if you use 'upmath' package.^^J}%
      }
  \else
  \fi
\fi

\ifCUPmtlplainloaded \else
  \checkfont{msam10}
  \iffontfound
    \IfFileExists{amssymb.sty}
      {\typeout{^^JFound AMS Symbol fonts on the system, using the
                'amssymb' package.^^J}%
       \usepackage{amssymb}%
         
         \let\geq=\geqslant
      }{}
  \fi
\fi

\ifCUPmtlplainloaded \else
  \IfFileExists{amsbsy.sty}
    {\typeout{^^JFound the 'amsbsy' package on the system, using it.^^J}%
     \usepackage{amsbsy}}
    {\providecommand\boldsymbol[1]{\mbox{\boldmath $##1$}}}
\fi

\newcommand{\Ztot}{\ensuremath{ Z\sub{tot}}}
\newcommand{\ntot}{\ensuremath{n\sub{tot}}}
\newcommand{\appref}[1]{\hyperref[#1]{Appendix~\ref{#1}}}

\newcommand{\blue}[1]{#1}
\newcommand{\sub}[1]{\ensuremath{_{\text{#1}}}}
\newcommand{\rd}{\ensuremath{\mathrm{d}}}

\usepackage{ae}
\usepackage{units}
\usepackage[disable]{todonotes}
\newcommand{\g}[1]{\mbox{\boldmath $#1$}}

\newcommand{\be}{\begin{displaymath}}
\newcommand{\ee}{\end{displaymath}}
\newcommand{\bn}{\begin{equation}}
\newcommand{\en}{\end{equation}}

\usepackage{enumerate}

\usepackage[T1]{fontenc}
\usepackage{bm}
\usepackage{color}
\usepackage{graphicx}
\usepackage[breaklinks=true]{hyperref}
\hypersetup{
  unicode=false,          % non-Latin characters in Acrobat’s bookmarks
  pdftoolbar=true,        % show Acrobat’s toolbar?
  pdfmenubar=true,        % show Acrobat’s menu?
  pdffitwindow=false,     % window fit to page when opened
  pdfstartview={FitH},    % fits the width of the page to the window
  pdfsubject={Subject},   % subject of the document
  pdfcreator={Creator},   % creator of the document
  pdfproducer={Producer}, % producer of the document
  pdfkeywords={keyword1} {key2} {key3}, % list of keywords
  pdfnewwindow=true,      % links in new PDF window
  colorlinks=true,        % false: boxed links; true: colored links
  linkcolor=blue,         % color of internal links (change box color with linkbordercolor)
  citecolor=blue,         % color of links to bibliography
  filecolor=blue,         % color of file links
  urlcolor=blue           % color of external links
}

\title{Dynamics of positrons during relativistic electron runaway}

\author{O.~Embr\'eus\aff{1}
  \corresp{\email{embreus@chalmers.se}},
  L.~Hesslow\aff{1}, M.~Hoppe\aff{1}, G.~Papp\aff{2}, K.~Richards\aff{1} \and T.~F\"ul\"op\aff{1}} 

\affiliation{\aff{1}Department of Physics, Chalmers University of Technology,
 SE-41296 G\"{o}teborg, Sweden
  \aff{2}Max-Planck-Institute for Plasma Physics, Garching, Germany}

\begin{document}
\maketitle
\begin{abstract}
  Sufficiently strong electric fields in plasmas can accelerate
  charged particles to relativistic energies. In this paper we
  describe the dynamics of positrons accelerated in such electric
  fields, and calculate the fraction of created positrons that become
  runaway accelerated, along with the amount of radiation that they
  emit. We derive an analytical formula that shows the relative
  importance of the different positron production processes, and show
  that above a certain threshold electric field the pair production by
  photons is lower than that by collisions. We furthermore present analytical and
  numerical solutions to the positron kinetic equation; these are
  applied to calculate the fraction of positrons that become
  accelerated or thermalized, which enters into rate equations that
  describe the evolution of the density of the slow and fast positron
  populations.  Finally, to indicate operational parameters required for
  positron detection during runaway in tokamak discharges, we give expressions for the parameter
  dependencies of detected annihilation radiation compared to bremsstrahlung
  detected at an angle perpendicular to the direction of runaway acceleration. Using the full 
  leading order pair production cross section, we demonstrate that previous 
  related work has overestimated the collisional pair production by at 
  least a factor of four.
\end{abstract}

\section{Introduction}
\label{sec:intro}
The production of positrons has been investigated extensively both
theoretically and experimentally since their first identification
\citep{Anderson238}.  Low energy positrons are used in many areas of
science and technology, ranging from positron emission tomography
\citep{Raichle1985} and surface science \citep{Hunt1999} to fundamental
studies of antimatter \citep{GabrielsePRL2002,Surko2004}. High energy
positrons can also be routinely produced in particle accelerators and
intense laser-solid interactions \citep{Chen2009,sarri_2015}.
Positrons are present in a wide range of atmospheric and astrophysical
plasmas, e.g.~lightning discharges \citep{Dwyer2014}, solar flares
\citep{Murphy2005}, pulsars and black-hole physics
\citep{Prantzos}. Also in post-disruption plasmas in large tokamaks,
where the energy of the runaway electrons is in the tens of MeV range,
high-energy positrons should be present \citep{helanderward,Fuloppapp},
but they have not yet been experimentally observed.

Plasmas with strong electric fields are particularly interesting for
positron generation, as particles accelerated by the field often reach
energies larger than the pair-production threshold. For example, the
electric field in solar flares is believed to be the result of
magnetic reconnection \citep{Priest2002,Liu2009}. In thunderstorms
strong electric fields are produced by the charged regions, sometimes
lasting tens of minutes \citep{Tsuchiya2011}.  In intense laser-matter
interaction, the positrons experience the sheath field that is set up
by the relativistic electrons leaving the target \citep{Wilks2001}. In
disruptive tokamak plasmas, the resistivity increase due to the sudden
cooling of the plasma leads to a high electric field that is induced
to maintain the plasma current \citep{Helander2002}. Regardless of the
cause, the electric field will strongly affect the dynamics of the
positrons.

If the electric field exceeds a certain critical field, the
accelerating force on the charged particles overcomes the friction,
and they are accelerated to high energies and run away
\citep{Wilson1925,Dreicer1959}. Existing runaway electrons may create
new (secondary) runaways in close-collisions with thermal electrons,
and this can lead to an exponential growth of the runaway population,
i.e.~an avalanche \citep{sokolov1979multiplication,Rosenbluthputvinski1997}.
%
%
% Note that ``gamma ray'' is the correct way to write it, EXCEPT for stellar gamma-rays, where it should be with a hyphen, as I left it now below.
The runaways are accelerated to energies that are well above the 
pair-production threshold \citep{hollmann15measurement,pazsoldan17spatiotemporal}  and create positrons in collisions with
electrons and ions. The created positrons are also accelerated by the electric
field, in the opposite direction with respect to the electrons, and if
the electric field is sufficiently strong, a substantial fraction of
them will run away \citep{Fuloppapp}. Eventually they will annihilate, either directly with electrons or through the formation of
positronium \citep{charlton2001positron}. Due to their drift motion, for runaway positrons in tokamaks this will typically occur after they have escaped the plasma and struck the first wall~\citep{liufisch2014}. 

The direct annihilation of an electron-positron pair at rest will
result in the creation of two gamma ray photons, each of energy 511~keV.  Positron annihilation is often invoked to explain the observed
emission features in the vicinity of 500~keV in the radiation spectrum
of gamma-ray bursts, pulsars, solar flares \citep{Murphy2005},
terrestrial lightning \citep{BriggsGRL} and the galactic centre
\citep{Prantzos}. In laboratory plasmas, the bremsstrahlung of the
energetic electrons may overwhelm the annihilation radiation from the
positrons, as the positron/electron fraction is usually
low \citep{Fuloppapp}. However, due to the directionality of the
bremsstrahlung radiation, the isotropic annihilation radiation may
still be detectable.

In this paper, we analyse the dynamics of high-energy positrons
produced in collisions between charged particles in a strong electric
field, where both electrons and positrons may run away. We use
MadGraph~5 simulations~\citep{madgraph} to obtain the cross section
for pair production in collisions between electrons and ions, which reveals that
the high-energy limit \citep{landauQED} or the formula 
given in \cite{gryaznykh1998} significantly overestimates
the cross section.  We consider the relative
importance of pair production by collisions and photons, and derive a
critical pair-production field above which collisional pair production
dominates in avalanching runaway scenarios.

In the case when pair production by photons is negligible, we solve
the kinetic equation for positrons. We derive an analytical expression
for the positron distribution function in the presence of an
avalanching runaway electron population.  The analytical results for
the distribution function and critical electric field are corroborated
with numerical simulations using the kinetic equation solver {\sc
  code} \citep{CODEpaper2014,Stahl2016} modified to include the
positron source and annihilation terms. Furthermore, we consider the
radiation emitted by positrons and find the parameter dependencies of
the annihilation to bremsstrahlung radiation ratio. This allows
determination of the parameter regions where the annihilation
radiation could be detectable in these plasmas.

The structure of the paper is the following. In Sec.~\ref{sec:source}
we describe the kinetic equation of the positrons including details of
the positron production source term.  We present both analytical and
numerical solutions of the kinetic equation, showing excellent
agreement in the relevant limit. Following this, in
Section~\ref{sec:rate} we describe rate equations for runaway
positrons, which are useful to predict the parametric dependencies of
the fraction of positrons without extensive kinetic simulations. In
Section~\ref{sec:rad} we calculate the expected annihilation radiation
from positrons in tokamak plasmas.  Finally we summarize our
conclusions in Sec.~\ref{sec:concl}.

\section{Kinetic equation for positrons}
\label{sec:source}
In this paper we consider the dynamics of positrons during a
relativistic electron runaway avalanche~\citep{jayakumar1993}. Due to the non-monotonic
dynamical friction acting on a charged test particle in a plasma, in
an electric field larger than a critical value $E_c$ fast electrons
may experience a net force that can rapidly accelerate them to
energies in the range of tens of MeV. In a fully ionized plasma, the
critical field is $E_c = \ln\Lambda n_e e^3 /
(4\pi\varepsilon_0^2 m_e c^2)$~\citep{connor}, where 
$\ln\Lambda \approx 14.6 +
0.5\ln(T[\text{eV}]/n_e[10^{20}\,\text{m}^{-3}])$ is the
Coulomb logarithm~\citep{solodovbetti}. We neglect a logarithmic energy dependence in
$\ln{\Lambda}$, and use the value for relativistic electrons at 1\,MeV
for simplicity.  Here, $n_e$ is the electron density, $e$ the
elementary charge, $\varepsilon_0$ the vacuum permittivity, $m_e$ the
electron rest mass and $c$ the speed of light.
The background plasma is assumed to be nearly Maxwellian for all species with the same temperature $T$.
In a neutral gas, $\ln{\Lambda}$
depends on the mean excitation energy of the medium instead of the
temperature, and corresponds to $\ln\Lambda \approx 11$ in
air~\citep{GurevichZybin2001}. In this case the electron density refers
to the density of bound electrons.

A sufficiently energetic electron can produce new runaway electrons
through elastic large-angle collisions. The result is an exponentially
growing number of runaway electrons, a so-called runaway
\emph{avalanche}. Each $e$-folding of the number density takes a time
$t\sub{ava} = c_Z / [4\pi n_e r_0^2 c (E/E_c - 1)]$ where $c_Z$ is
only weakly dependent on electric field, and can be approximated by
$c_Z \approx \sqrt{5+Z\sub{eff}}$ in a fully ionized
plasma~\citep{Rosenbluthputvinski1997}, where the effective charge is
$Z\sub{eff} = \sum n_i Z_i^2 / \sum_i n_i Z_i$ with the sum taken over
all ion species $i$. 
%LINNEA ta bort?
We shall find that several results in the paper
are insensitive to the details of $c_Z$, assuming only that it is
independent of $E$. As such, more accurate models of the avalanche
process can in principle be implemented by inserting for $c_Z$ the value characterizing any particular scenario of interest.

Since the electrons are ultra-relativistic, they will create positrons
which are predominantly co-moving; these are created either directly
in collisions or indirectly through the hard X-rays emitted in 
collisions~\citep{heitler}, which can produce a pair in a subsequent
interaction. Since the positrons experience an acceleration by the
electric field in the direction opposite to the runaway-electron
motion, they will immediately start decelerating. A fraction of these
positrons will slow down to thermal speeds where they eventually annihilate,
whereas the remainder obtain sufficiently large momenta perpendicular
to the acceleration direction that they become runaway-accelerated
along the electric field, moving anti-parallel to the runaway
electrons. Annihilation -- which occurs at a rate that decreases with
positron energy -- does not have a significant effect on the dynamics of the
energetic positrons since the avalanche rate is typically much faster,
which is demonstrated in Sec.~\ref{sec:distribution}. 

Throughout this paper, we shall assume that the plasma is fully
ionized. In a partially ionized plasma or a neutral gas, screening
effects due to the bound electrons would enter into all binary
interactions. In the 10 MeV energy range, these are however largely
negligible for the pair production mechanisms as well as for the
emission of bremsstrahlung, meaning that they are to be
calculated using the full nuclear charge of the target. The screening
effects become significant when $p/m_e c\gtrsim 137/Z^{1/3}$~\citep{heitler}. 
Elastic Coulomb collisions are to a greater extent affected by screening
effects, where the pitch-angle scattering rates may be reduced by
approximately up to two thirds and energy-loss rates by one
third~\citep{Hesslow} in the energy range of interest, compared to 
the results obtained treating the
medium as fully ionized. This would modify primarily two important
quantities that affect our results: the avalanche growth rate factor
$c_Z$, as well as the critical field $E_c$ \citep{Hesslow2018}, which can here be assumed to be
accurate only up to an order-of-unity factor in partially ionized plasmas.
While the results we present are strictly valid for a fully ionized plasma,
we expect to capture the correct order of magnitude also in a partially 
ionized plasma or neutral gas, if the effective charge and electron densities
appearing in the formulas are always evaluated using the fully-ionized values. We denote these by
\begin{align*}
\ntot &= \sum_i Z_i n_i, \\ %LINNEA ha n\sub{tot} i resten
\Ztot &= \frac{1}{\ntot}\sum_i n_i Z_i^2,
\end{align*}
where $Z_i$ is the atomic number of species $i$. Thus, the density is always to be taken as the total density of free plus bound electrons, and in a single-component gas or plasma $Z\sub{tot}$ is the atomic number of the ion species regardless of ionization degree. 

The dynamics described above can be most lucidly captured in a
two-dimensional model.  The distribution function of positrons with momentum $\g{p} = m_e \g{v}/\sqrt{1-v^2/c^2}$, where the positron velocity  is denoted \g{v}, at a time $t$ is denoted
$f\sub{pos}(t,\,\g{p})$. In a homogeneous cylindrically symmetric
plasma in the presence of an electric field \g{E} it satisfies the
kinetic equation
\begin{align}
\frac{\partial f\sub{pos}}{\partial t} + eE
\left[\xi\frac{\partial}{\partial p} +
  \frac{1-\xi^2}{p}\frac{\partial}{\partial \xi}\right]f\sub{pos} =
C\sub{pos} + S\sub{pos} +S\sub{an},
\label{eq:FPeq}
\end{align}
where $E=|\g{E}|$, $p=|\g{p}|$,
$\xi \equiv \cos\theta = \g{p}\cdot \g{E}/pE$ is the pitch-angle
cosine, $C\sub{pos}$ is the positron collision operator, $S\sub{pos}$
denotes the source term of positrons generated in collisions between
relativistic runaway electrons and field particles of the plasma, as
well as positron production by highly energetic photons, and
$S\sub{an}$ denotes the annihilation term. In a magnetized plasma, the
equation is valid for an axisymmetric positron distribution if $E$ is replaced by the component of the electric
field parallel to the magnetic field, and the pitch angle is instead
defined relative to the magnetic field.

In the limit of small energy transfers, the elastic positron-electron
and positron-ion differential scattering cross sections coincide with the
electron-electron and electron-ion cross sections,
respectively~\citep{landauQED}. Consequently, the positron collision
operator $C\sub{pos}$ equals the electron collision operator $C_e$ up
to terms small in the Coulomb logarithm $\ln\Lambda$. Large-angle collisions,
which are primarily important for avalanche generation when $\ln\Lambda$ is large, can be neglected since the thermal positron population will always be small in number. The positron
distribution therefore satisfies the same kinetic equation as the
electron distribution, except for the electric field accelerating them
in the opposite direction (with these definitions positrons are
accelerated towards $\xi=1$, and electrons towards $\xi=-1$), and the
presence of the terms $S\sub{pos}$ and $S\sub{an}$ describing
their creation and annihilation, respectively.

%Runaway electrons produce electron-positron pairs by two main
%mechanisms: in collisions between runaway electrons with energy above
%$3m_e c^2$ and the cold background, as well as by photons emitted %%Geri: Do we have to cite the 3 m_e c^2? Intersting to note that pair production in the vicinity of an electron requires 4mec^2 energy.
%through bremsstrahlung with energy larger than $2m_ec^2$ in a subsequent interaction with the background. 
The number of positrons created with momentum \g{p} in time $\rd t$ has two main
contributions: (\emph{1}) the collisions between stationary ions of
species $i$ with density $n_i$ and the $\rd n\sub{RE}$ number of
runaways at momentum $\g{p}_1$ and speed  $v_1$, and (\emph{2}) the pair production of
$\rd n_\gamma$ photons in the field of ions $i$:
\begin{align}
\rd n\sub{pos} = \sum_i \Big[n_i v_1 \rd \sigma^+_{ci} \rd n\sub{RE} \rd t + n_i c \rd \sigma^+_{\gamma i} \rd n_\gamma \rd t\Big]. 
\end{align}
Here, $\rd\sigma^+_{ci}$ is the differential cross section for
producing a positron in a collision between an electron and a
stationary ion, and similarly $\rd \sigma^+_{\gamma i}$ for a photon
interacting with stationary ions, and are given in
Appendix~\ref{ap:source}.  We use the Madgraph~5 tool~\citep{madgraph}
for obtaining the pair-production cross sections throughout this
paper. 

Using $\rd n\sub{RE}(\g{p}_1) = f\sub{RE}(\g{p}_1) \rd \g{p}_1$, where
$f\sub{RE}$ is the distribution function of runaway electrons, and
similarly for the positron distribution
$f\sub{pos}(\g{p}) = \rd n\sub{pos}/\rd \g{p}$ and photons
$\phi_\gamma(\g{k}) = \rd n_\gamma/\rd\g{k}$ where $\g{k}/c$ is the
photon momentum, we find the following form for the positron source
$S\sub{pos}$:
\begin{align}
S\sub{pos}\equiv\left(\frac{\partial f\sub{pos}}{\partial
  t}\right)\sub{pp} = \sum_i n_i c\Biggr[\int\rd \g{p}_1 \,\frac{v_1}{c} \frac{\partial
  \sigma^+_{ci}}{\partial \g{p}} f\sub{RE}(\g{p}_1)
   +  \int \rd \g{k} \, \frac{\partial \sigma_{\gamma i}^+}{\partial \g{p}} \phi_\gamma(\g{k})\Biggr]. 
\end{align}
In an avalanching runaway scenario, the photon distribution can be
eliminated \blue{in favour of an expression involving only the runaway distribution} because of the relatively slow evolution of \blue{the photon} energy spectrum. The
runaway-electron population grows exponentially in time on the
time-scale~\citep{Rosenbluthputvinski1997}
$$t\sub{ava} = \frac{c_Z}{4\pi \ntot r_0^2 c(E/E_c-1)}. $$ 
The photons on the other hand evolve on the Compton-scattering time
scale~\citep{heitler}
$$t\sub{Co} = \frac{k}{\pi \ntot r_0^2 c m_e c^2 \ln\left[2k/(m_e c^2)\right]},$$ 
where the photon energies $k = |\g{k}|$ are larger than the 
pair-production threshold $2m_e c^2$, and
$r_0 = e^2/(4\pi\varepsilon_0 m_e c^2) \approx 2.82\cdot 10^{-15}$\,m
is the classical electron radius.  Comparing the two time scales shows
that the photons do not have time to change significantly from the distribution in
which they are created whenever
$$\frac{k/m_e c^2}{\ln(2k/m_e c^2)}\gg \frac{c_Z}{4(E/E_c-1)}.$$ Since
the right-hand side is typically smaller than unity, this is
generally well satisfied in an avalanching runaway
scenario. The photon distribution is then given by
\begin{align}
\phi(\g{k}) = t\sub{ava}\sum_i n_i \int \rd \g{p}_1 \,v_1 \frac{\partial \sigma\sub{br,$i$}}{\partial \g{k}}(\g{k},\,\g{p}_1) f\sub{RE}(\g{p}_1),
\end{align}
where $\rd \sigma\sub{br,$i$}$ is the differential bremsstrahlung
cross section for interactions between electrons and particle species
$i$.

Since the cross sections appearing in these formulas depend on target
species only through $Z_i^2$, the target charge squared~\citep{heitler},  
these may be factored out when screening effects are neglected, yielding a factor of the
effective plasma charge $\Ztot $ when summed over $i$.  We shall therefore suppress the 
indices $i$ of the cross sections by writing
$\sum_i n_i \sigma_{ci}^+ = \ntot \Ztot  \sigma_c^+$ for collisional pair production, and 
$\sum_i n_i \sigma\sub{br,$i$} = \ntot (\Ztot+1)\sigma\sub{br}$ (and likewise for $\sigma_\gamma^+$) for the photon pair production cross sections. Here we have added the contribution from electron-electron bremsstrahlung in the approximation that the electron-electron and electron-proton bremsstrahlung cross sections are the same, which has satisfactory accuracy since the majority of interactions occur with negligible momentum transfer to the target particle~\citep{haug1975}. Conversely, for collisional pair production the electron-electron interactions are negligible, which was verified with MadGraph~5 simulations~\citep{madgraph,gabriele} which indicated that the $e$-$e$ cross section is 10-20\% of the $e$-$i$ cross section when the incident electron lab-frame energy ranges over 10-20 MeV and $Z_i=1$.

The positron source can then be written
\begin{align}
S\sub{pos} = \Ztot  \ntot  \int \rd \g{p}_1 \, v_1 \frac{\partial \sigma^+}{\partial \g{p}}f\sub{RE}(\g{p}_1),
\end{align}
where the effective pair-production cross section $\rd \sigma^+$,
accounting for both direct pair production in collisions as well as by
X-rays, is given by
\begin{align}
\frac{\partial \sigma^+}{\partial \g{p}} = \frac{\partial \sigma_{c}^+}{\partial \g{p}} + \frac{(\Ztot +1)^2}{\Ztot} t\sub{ava} \ntot c  
 \int \rd \g{k} \, \frac{\partial \sigma_\gamma^+}{\partial \g{p}}(\g{p},\,\g{k}) \frac{\partial \sigma\sub{br}}{\partial \g{k}}(\g{k},\,\g{p}_1).
\end{align}

Positrons are created with a significant fraction of the energy of the
incident electron that created them, but with a momentum perpendicular
to the direction of the incident electron of 
order~\citep{landauQED,heitler} $p_\perp \approx m_e c$. This means
that the differential cross section for their production is strongly
peaked in the direction of the incident electron; throughout this work
we assume that it is delta distributed in the scattering angles, and write
\begin{align}
\frac{\partial \sigma^+_{c}}{\partial \g{p}} &= \frac{\delta(\cos\theta-\cos\theta_1) }{2\pi (m_e c)^2 p \gamma}\frac{\partial \sigma^+_{c}}{\partial \gamma}(p,\,p_1),\nonumber \\
\frac{\partial \sigma_\gamma^+}{\partial \g{p}} &= \frac{\delta(\cos\theta-\cos\theta_k) }{2\pi (m_e c)^2 p \gamma}\frac{\partial \sigma^+_\gamma}{\partial \gamma}(p,\,k), \label{eq:delta cs} \\
\frac{\partial \sigma\sub{br}}{\partial \g{k}} &= \frac{\delta(\cos \theta_k - \cos\theta_1)}{2\pi k^2} \frac{\partial \sigma\sub{br}}{\partial k}(k,\,p_1), \nonumber
\end{align}
where $\gamma = \sqrt{1+(p/m_e c)^2}$ is the Lorentz factor. The angles $\theta$, $\theta_1$ and $\theta_k$ are the angles between the accelerating electric field $\g{E}$ (or in a magnetized plasma the magnetic field $\g{B}$) and $\g{p}$, $\g{p}_1$ and $\g{k}$, respectively.
With an axisymmetric runaway distribution
$f\sub{RE}(\g{p}_1) = f\sub{RE}(p_1,\,\cos\theta_1)$, we then obtain
the approximated positron source term
\begin{align}
S\sub{pos}(\gamma,\cos\theta) = \frac{\ntot \Ztot  m_e c^2}{p\gamma}\int_{\gamma+2}^\infty \hspace{-1mm} \rd \gamma_1 \, (\gamma_1^2-1) {\frac{\partial \sigma}{\partial \gamma} \hspace{-0.8mm}}^+ \hspace{-1mm}f\sub{RE}(\gamma_1,\cos\theta),
\label{eq:full source}
\end{align}
where the effective cross section now takes the form
\begin{align}
\frac{\partial \sigma^+}{\partial \gamma}(\gamma,\,\gamma_1) = \frac{\partial \sigma^+_c}{\partial \gamma}(\gamma,\,\gamma_1)+ \frac{(\Ztot+1)^2}{\Ztot} t\sub{ava}\ntot c 
\int_{\gamma+1}^{\gamma_1-1} \rd k \,\frac{\partial \sigma_\gamma^+}{\partial \gamma}(\gamma,\,k)\frac{\partial \sigma\sub{br}}{\partial k}(k,\,\gamma_1),\label{eq:full cs}
\end{align}
depending only on the simpler integrated cross sections that are only
differential in the energy of the outgoing particle of interest. In
Eq.~(\ref{eq:full source}) it is explicit that positrons are only
generated in the direction of the incident energetic electrons -- the
electron distribution is sampled at the same pitch angle as the
source. Further details of the positron source term in the presence of
an avalanching electron distribution are given in \appref{ap:source},
where we present the differential cross sections used as well as
illustrate typical shapes of the source term in Eq.~(\ref{eq:full source}).

The annihilation source takes the simpler form
\begin{align}
S\sub{an}(\bold{p}) = -\ntot v \sigma\sub{an}(p)f\sub{pos}(\bold{p}),
\end{align}
where $\sigma\sub{an}$ is the cross section for free positron-electron two-quanta annihilation against stationary target electrons~\citep{heitler}
\begin{align}
\sigma\sub{an} =   \frac{\pi r_0^2}{(p/m_e c)(\gamma+1)}\left[ \frac{\gamma^2+4\gamma+1}{p/m_e c}\ln\left(\gamma+\frac{p}{m_ec}\right) - \gamma-3\right].
\label{eq:annihilation cs}
\end{align}

\subsection{Relative importance of pair production by collisions and by photons} 
A peculiar phenomenon occurs when considering pair production in the
presence of a strong electric field, where the number of energetic
electrons grows exponentially in time. Because there is a delay
between the emission of photons and their subsequent pair production,
if the electron population has time to grow by a significant amount
during one such photon-pair-production time, the direct positron
generation in collisions may contribute with a relatively larger
production of pairs. We will now proceed to
derive the threshold electric field above which pair production in
collisions is dominant due to this effect.

In order to evaluate the pair-production source terms we need an
expression for the runaway electron distribution. In a spatially
uniform fully ionized plasma with constant electric field, when the runaway generation is dominated by the avalanche mechanism -- i.e.~by multiplication through large-angle collisions -- it is given
by~\citep{Fulop2006} 
\begin{align}
f\sub{RE}(p,\,\xi,\,t) &=   \frac{n\sub{RE}(t)A(p) }{2\pi m_e c\gamma_0  p^2 } \frac{\exp\left[-\frac{\gamma}{\gamma_0}-A(p)(1+\xi)\right]}{1-e^{-2A}},
\label{eq:full RE dist} \\
A(p) &= \frac{E/E_c+1}{\Ztot +1}\gamma, \nonumber \\
n\sub{RE}(t) &= n\sub{RE}(0) e^{t/t\sub{ava}},\nonumber \\
\gamma_0 &= c_Z\ln\Lambda. \nonumber 
\end{align}
Our choice for $A$ differs slightly from that in
Ref.~\citep{Fulop2006}, however agrees in the limit $E \gg E_c$,
$p\gg m_e c$ and $1+\xi \ll 1$ where the solution is expected to be
valid, but is here generalized to also capture the near-threshold
limit $E \to E_c$~\citep{lehtinen,Hesslow2018}. When pitch-angle
averaged, the electron distribution is given by
\begin{align}
\hspace{-1mm}\mathcal{F}\sub{RE}(p,t) = 2\pi\int_{-1}^1\rd \xi \, f\sub{RE}(p,\xi,t) =  \frac{n\sub{RE}(t)}{m_e c \gamma_0}e^{-\gamma/\gamma_0},
\label{eq:1D RE dist}
\end{align}
where the average runaway energy is given by
$\gamma_0 m_e c^2 \approx (c_Z\ln\Lambda/2)$\,MeV, which is typically of the order of 10-30\,MeV in most scenarios of interest.

The total number of pairs created per unit time and volume is obtained
by integrating the positron source function (\ref{eq:full source})
over all momenta, yielding
\begin{align}
\frac{\rd n\sub{pair}}{\rd t}  %n_e \Ztot m_e c^2\int_1^\infty \rd \gamma \int_{\gamma+2}^\infty \rd \gamma_1 \, \frac{\partial \sigma^+}{\partial \gamma}\mathcal{F}\sub{RE}(\gamma_1) \label{eq:total production rate} \nonumber \\
&= n_e \Ztot  m_e c^2\int_3^\infty \rd \gamma_1 \,\sigma^+(\gamma_1) \mathcal{F}\sub{RE}(\gamma_1)\equiv  n_e n\sub{RE} \Ztot  \langle v\sigma^+ \rangle\sub{RE} , \\
\sigma^+(\gamma_1) &= \int_1^{\gamma_1-2} \frac{\partial \sigma^+}{\partial \gamma}(\gamma,\,\gamma_1)\,\rd \gamma, \nonumber \\
\langle v\sigma^+ \rangle\sub{RE} &= \frac{1}{n\sub{RE}}\int_{\sqrt{8}}^\infty \rd p_1 \, v_1\sigma^+(\gamma_1) \mathcal{F}\sub{RE}(\gamma_1). 
\end{align}
With the analytic form of Eq.~(\ref{eq:1D RE dist}) for the electron distribution, the pair production rate defined by the above equations is characterized by the two integrals 
\begin{align}
\langle v \sigma_c^+ \rangle\sub{RE} &= \frac{m_e c^2}{n\sub{RE}} \int_3^\infty \rd \gamma_1 \,\sigma_c^+(\gamma_1)\mathcal{F}\sub{RE}(\gamma_1) \approx \alpha^2  r_0^2 c  \frac{\gamma_0-6.7}{15}, \nonumber\\
\langle v \sigma_\gamma \rangle\sub{RE}& = \frac{m_e c^2}{n\sub{RE}} \int_3^\infty \rd \gamma_1 \, \mathcal{F}\sub{RE}(\gamma_1) \int_2^{\gamma_1-1}\rd k \, \sigma_\gamma^+(k)\frac{\partial \sigma\sub{br}}{\partial k}(k,\,\gamma_1)  
\approx \alpha^2 r_0^4 c \big( 2.6\gamma_0 - 14.8 \big),
\end{align}
where $\sigma_\gamma^+ = \int_1^{k-1}(\partial \sigma_\gamma^+/\partial
\gamma)\,\rd \gamma $, and the approximate formulas are least-square fits on the interval of $\gamma_0$ between 20 and 80, giving a maximum error of 2.5\%. Within an error of less than $3\%$, the second expression differs from the first by a constant factor $40.75 r_0^2$, allowing the total pair production rate to be written
\begin{align}
\frac{\rd n\sub{pair}}{\rd t} \approx Z\sub{tot}\alpha^2 n_e r_0^2 c \frac{\gamma_0-6.7}{15}
\left( 1 + 40.75\frac{(Z\sub{tot}+1)^2}{Z\sub{tot}} t\sub{ava}n_e c r_0^2 \right).
\end{align}
With $t\sub{ava} n_e c r_0^2 = c_Z/[4\pi(E/E_c - 1)]$, it is clear that there is a threshold field $E=E\sub{pp}(Z\sub{tot})$ above which the collisional pair production (described by the first term) will be dominant. When $c_Z$ is independent of $E$, \blue{this threshold field} is given by
\begin{align}
\frac{E\sub{pp}}{E_c}-1 &= \frac{(1+Z\sub{tot})^2}{Z\sub{tot}}\frac{c_Z}{4\pi r_0^2} 
\frac{\int_3^\infty \rd \gamma_1\,e^{-\gamma_1/\gamma_0}\int_2^{\gamma_1-1} \rd k \,\sigma_\gamma^+(k)\frac{\partial \sigma\sub{br}}{\partial k}(k,\,\gamma_1)}{\int_3^\infty\rd \gamma_1 \,e^{-\gamma_1/\gamma_0}\sigma_c^+(\gamma_1)} \nonumber\\
&\approx 3.25 c_Z\frac{(1+Z\sub{tot})^2}{Z\sub{tot}}.
\label{eq:threshold field}
\end{align}

%The contributions from the two pair-production mechanisms will be
%equally large when
%\begin{align}
%&\frac{4\pi r_0^2 \Ztot}{(1+\Ztot)^2 c_Z}\left(\frac{E}{E_c}-1\right)\int_3^\infty \rd \gamma_1 \,e^{-\gamma_1/\gamma_0}\sigma_c^+(\gamma_1) \nonumber \\
%&=\int_3^\infty \rd \gamma_1\,e^{-\gamma_1/\gamma_0}\int_2^{\gamma_1-1} \rd k \,\sigma_\gamma^+(k)\frac{\partial \sigma\sub{br}}{\partial k}(k,\,\gamma_1),
%\label{eq:Epp general eq}
%\end{align}
%where $\sigma_c^+$ is defined in a manner analogous to $\sigma^+$
%above, and
%$\sigma_\gamma^+ = \int_1^{k-1}(\partial \sigma_\gamma^+/\partial
%\gamma)\,\rd \gamma $.
%Note, that a charge dependence also enters through $\gamma_0$ in the
%integrals. From this equation we can express the critical
%pair-production field $E = E\sub{pp}(\Ztot )$ above which
%collisional pair production dominates, and which is plotted in
%Fig.~\ref{fig:critical pp field} as a function of $\Ztot $ for two
%different values of $\ln\Lambda$ using $c_Z = \sqrt{\Ztot +5}$.  
%The integrals appearing in the
%expression, which depend only on the average runaway energy
%$\gamma_0 = c_Z\ln\Lambda$, can be well captured by a simple numerical
%fit yielding the convenient expression for the threshold field
%\begin{align}
%\frac{E\sub{pp}}{E_c} - 1 \approx  c_Z \frac{(1+\Ztot)^2}{\Ztot}\left(1-\frac{49}{59+c_Z\ln\Lambda}\right), \end{align}
%with a relative error compared to Eq.~(\ref{eq:Epp general eq}) less
%than $1\%$ when $20 < \gamma_0 < 180$.

When $\Ztot  = 1$, the threshold field is
$E\sub{pp}\approx 33E_c$, but grows
rapidly with $\Ztot $. With an air-like $\Ztot  = 8$, one obtains $E\sub{pp}\approx 120E_c$. Since
electric fields are typically close to threshold during lightning
discharges, positron production in such scenarios can be expected to
be dominated by photon pair production. \blue{Although this has been assumed to be true in previous atmospheric runaway studies, the domain of validity of such an assumption has not been discussed}.

In the above we assumed an infinitely large homogeneous system. When
runaway acceleration occurs only over a finite distance of length $L$
of constant background parameters, the threshold field calculated
above is valid when $L \gg L\sub{ava} = c t\sub{ava}$, that is, when
the system is significantly longer than one avalanche mean-free
path. When this is not satisfied, i.e. when $L\lesssim L\sub{ava}$, a
threshold condition for the length of the system is obtained instead,
taking the form
\begin{align}
L\sub{pp} \approx \frac{\Ztot}{(1+\Ztot)^2}\frac{3\times 10^7\,\text{m}}{ n_e[10^{20}\,\text{m}^{-3}]}.
\end{align}
When $L\sub{pp} \lesssim L \lesssim L\sub{ava}$, photon pair production will be the dominant positron-generation mechanism.

\begin{figure}
\begin{center}
\includegraphics[width=0.6\textwidth,trim=7mm 0 9mm 0]{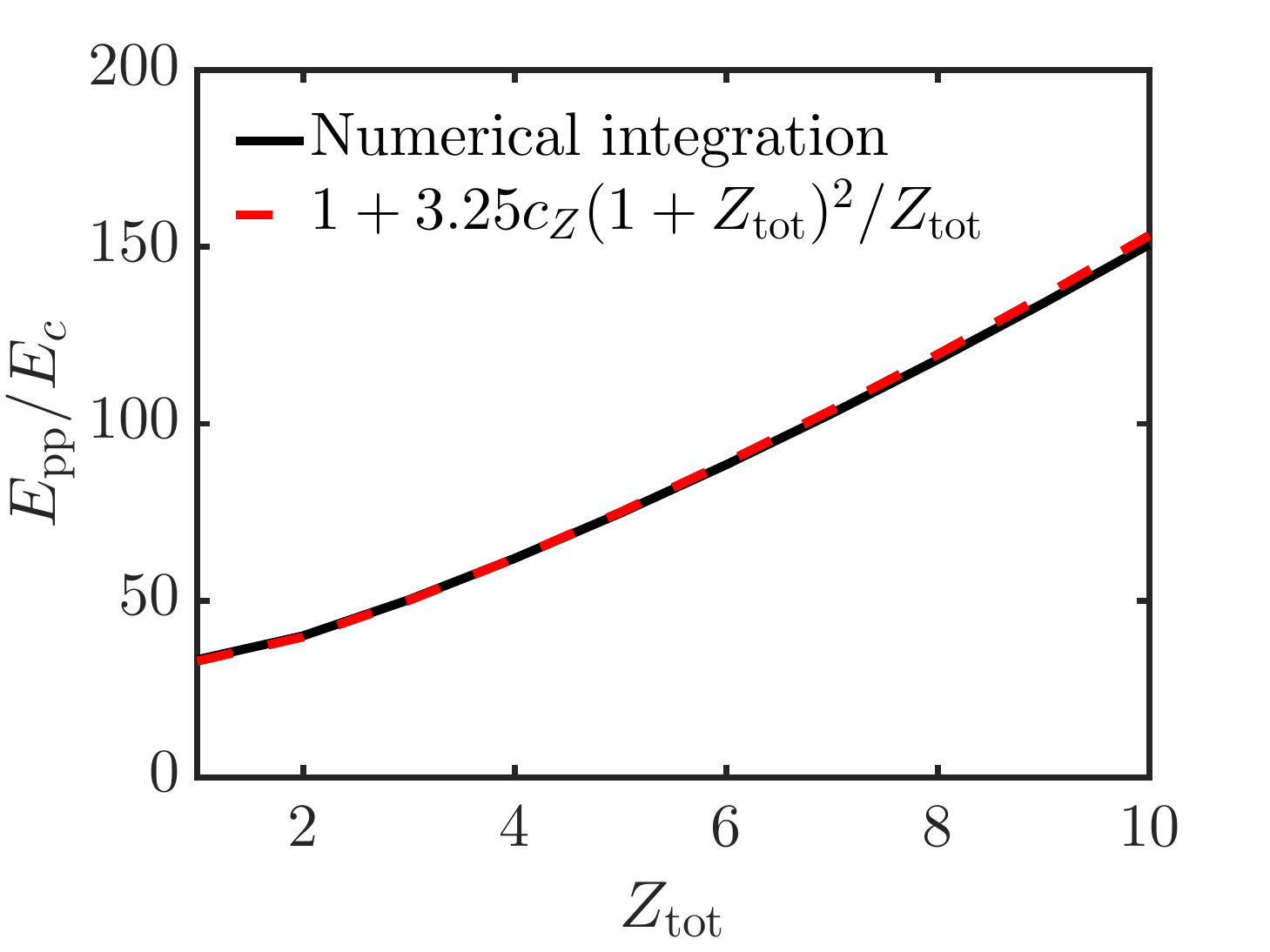}
\caption{Critical field $E\sub{pp}$ above which collisional positron production is the dominant pair-production mechanism in a uniform plasma, normalized to the avalanche threshold field $E_c$, calculated from the two expressions given in in Eq.~(\ref{eq:threshold field}) with $c_Z = \sqrt{5+Z\sub{tot}}$. \label{fig:critical pp field}}
\end{center} 
\end{figure}

In the remainder of this work, we will focus on scenarios where either
$E \gtrsim E\sub{pp}$ or $L \lesssim L\sub{pp}$, so that pair
production by photons is negligible. This is typical of runaway
scenarios in tokamaks, where $L/L\sub{pp} \ll 10^{-5}$ due to the
small size of the device.

\subsection{Distribution function of fast positrons}
\label{sec:distribution}
Equipped with the kinetic equation for positrons in a runaway
scenario, we can now characterize its solutions.  When the electric
field is sufficiently large for the average pitch angle to be small,
typically well satisfied when
$A = (E/E_c+1)\gamma/(\Ztot +1) \gtrsim 1$, the distribution
function of fast positrons can be readily calculated analytically. The
kinetics are then essentially one-dimensional, with the pitch-angle
dynamics playing a peripheral role in the evolution of the energy
spectrum.

We introduce a half-plane pitch-angle-averaged positron distribution function 
$\mathcal{F}$ as
\begin{align}
\mathcal{F}(p) =2\pi p^2\times \begin{cases}
\int_0^1 \rd \xi \,f\sub{RP}(p,\,\xi), & p \geq 0 \\  
\int_{-1}^0 \rd \xi \,f\sub{RP}(|p|,\,\xi), & p < 0
\end{cases}
\end{align}
where the coordinate $p$ now ranges from $-\infty$ to $\infty$.
This distribution is defined so that $\int_{p_c}^\infty \mathcal{F} \, \rd p = n\sub{RP}$ equals the total runaway-positron density, with $p_c$ a superthermal threshold in momentum distinguishing thermal positrons from runaways. In the same way, the thermal number density of positrons is $n\sub{TP} = \int_{-p_c}^{p_c}\mathcal{F}\,\rd p$.

In Appendix~\ref{ap:distributions} we solve the positron kinetic equation (\ref{eq:FPeq}) in the limit $(p/m_e c)^2 \gg 1$ assuming small pitch-angles $1-|\xi|\ll1$. The resulting positron distribution is given by
\begin{align}
\mathcal{F}(p) = \frac{\Ztot }{4\pi\ln\Lambda r_0^2}\frac{n\sub{RE}(t)}{\gamma_0 m_e c} e^{\rho\gamma/\gamma_0}  
 \int_\gamma^\infty \rd \gamma' \int_{\gamma'+2}^\infty \rd \gamma_1 \, \frac{\partial \sigma^+}{\partial \gamma'}(\gamma',\gamma_1)\exp\left(-\frac{\rho\gamma'+\gamma_1}{\gamma_0}\right)
\label{eq:ss dist}
\end{align}
for $p<0$, which describes the slowing-down distribution of the newly created positrons, where $\rho = ({E/E_c-1})/({E/E_c+1})$, and

\begin{align}
\mathcal{F}(p,\,t) &= \frac{n\sub{RP}(t)}{m_e c \gamma_0} e^{-\gamma/\gamma_0}
\label{eq:rp dist}, \\
n\sub{RP}(t) &= n\sub{RP}(0)e^{t/t\sub{ava}} \nonumber
\end{align}
for $p>0$, describing the runaway positron population that is undergoing acceleration in the electric field. Note that the prefactor, including the runaway-positron density evaluated at $t=0$, is not determined in this derivation, but must instead be calculated in a more comprehensive kinetic equation accounting for the dynamics near $p \lesssim m_e c$. 

We see that when $p>0$, the runaway positron distribution satisfies $\mathcal{F}(p) = ({n\sub{RP}}/{n\sub{RE}})\mathcal{F}\sub{RE}(-p)$. 
Indeed, for the full positron distribution, since the kinetic equation (\ref{eq:FPeq}) is identical to the runaway-electron equation for $\xi>0$ where the pair-production source vanishes, we would expect 
\begin{align}
f\sub{RP}(p,\,\xi,\,t) \approx \frac{n\sub{RP}(t)}{n\sub{RE}(t)} f\sub{RE}(p,\,-\xi,\,t).
\end{align}
The expressions given above are valid for collisional as well as for photon pair-production during runaway scenarios. 

%\paragraph{Runaway annihilation time}
We can now accurately evaluate the annihilation rate of runaway positrons, obtaining in the ultra-relativistic limit,
\begin{align} 
\frac{1}{\tau\sub{aR}} = \frac{n_e }{n\sub{RP}}\int_{p_c}^\infty \rd p \, v \mathcal{F}(p) \sigma\sub{an}(p)   
&\approx \frac{1}{4 \ln\Lambda(E/E_c-1) t\sub{ava}}\int_1^\infty \rd \gamma \frac{\ln 2\gamma-1}{\gamma}e^{-\gamma/\gamma_0} \nonumber \\
%&\sim \frac{ \ln^2(\gamma_0/\gamma_e)  - 2(1-\ln 2)\ln(\gamma_0/\gamma_e)+ \pi^2/6}{8\ln\Lambda\left({E}/{E_c}-1\right) t\sub{ava}}\nonumber \\
&\approx \frac{ \ln^2(\gamma_0/2.42)+1.55 }{8\ln\Lambda (E/E_c-1) t\sub{ava}},
\end{align}
the final approximation having an error less than $2\%$ for $\gamma_0>20$, and where the annihilation cross section $\sigma\sub{an}$ was given in equation (\ref{eq:annihilation cs}).  We find that typically $t\sub{ava}/\tau\sub{aR} \lesssim 0.1/(E/E_c -1)$, showing that annihilation has negligible impact on the avalanche-time-scale dynamics except for very close to the threshold field $E_c$. At that point, however, most of the created positrons will become thermalized, and only a negligible fraction will have time to annihilate before reaching thermal energies.

For $v\ll c$ the annihilation cross section takes the simple form $\sigma\sub{an}\sim \pi r_0^2 c/v$, so that the thermal annihilation time $\tau\sub{aT}$ for a thermal positron population of temperature $T \ll 511\,$keV is given simply by
\begin{align}
\frac{1}{\tau\sub{aT}} = \pi n_e r_0^2 c = \frac{1}{4t\sub{ava}}\frac{c_Z}{E/E_c - 1}.
\end{align}
In the presence of partially ionized or neutral gases, however, the cold positrons may annihilate also through the formation of positronium, which has significantly shorter, sub-$\mu$s lifetime. The annihilation time of thermal positrons is then rather set by the positronium formation rate, which is of the order of $n_i v a_0^2$, with $a_0$ the Bohr radius~\citep{charlton2001positron}.
%typically satisfying $\tau\sub{aT} \ll t\sub{ava}$. 

\subsection{Numerical distribution function} The positron
Fokker-Planck equation, Eq.~(\ref{eq:FPeq}), can be solved as an
initial value problem to give the evolution of the positron
distribution function in the presence of an accelerating electric
field. By adding the source and annihilation terms to the {\sc code}
\citep{CODEpaper2014,Stahl2016} numerical kinetic solver, we calculate the
distribution function for various electric fields and effective ion
charges.  {\sc code} uses a continuum-spectral discretization scheme and has
been used extensively to calculate runaway electron distributions
including partial screening effects \citep{Hesslow}, synchrotron
radiation \citep{Stahl2015,hirvijoki_2015}, bremsstrahlung
\citep{EmbreusBrems2016}, and close collisions \citep{olaknockon2018}.

\begin{figure}
  \begin{center}
    \includegraphics[width=0.6\textwidth,trim=7mm 0 5mm 0]{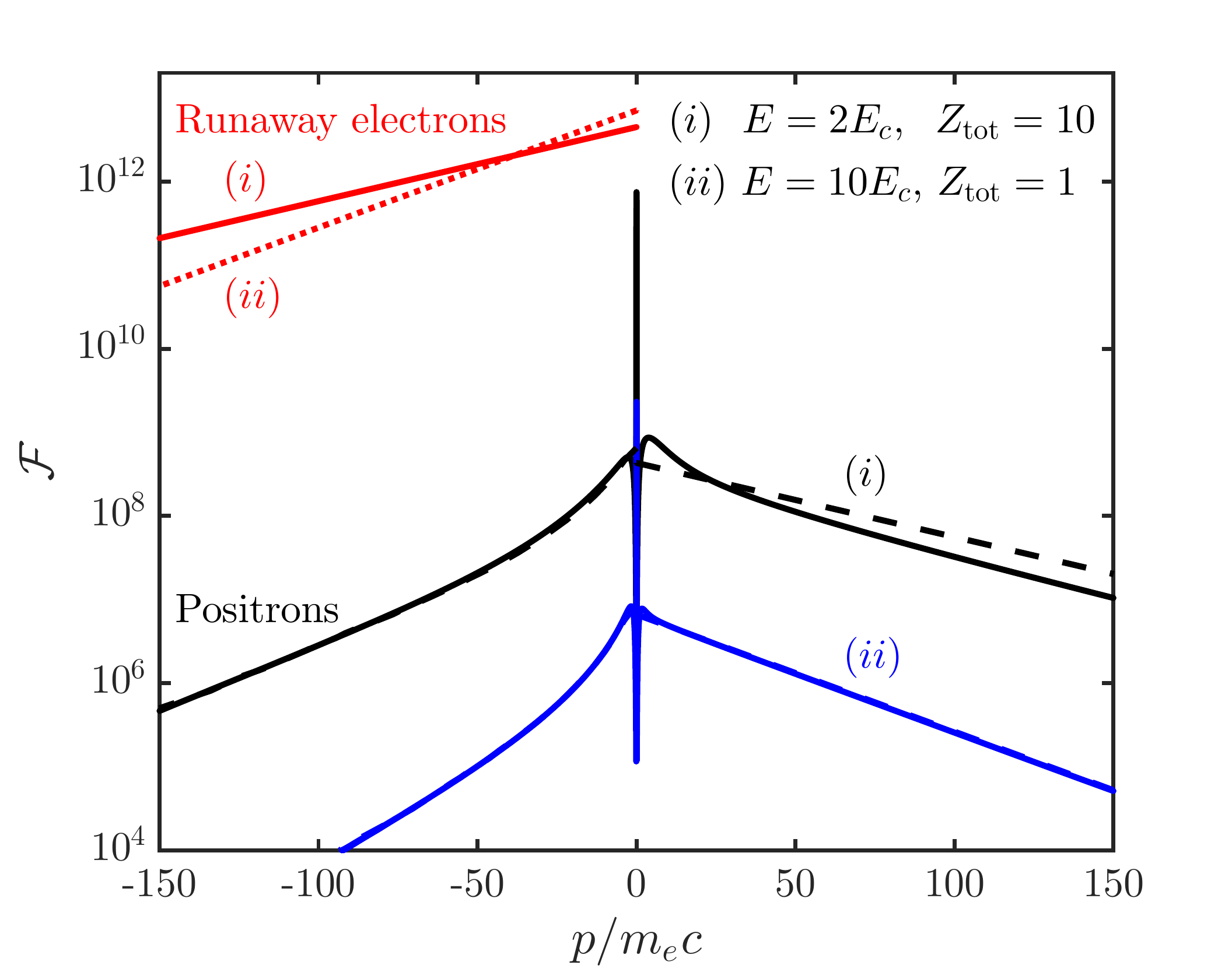}
    \end{center}
\caption{Pitch-angle averaged distribution functions $\mathcal{F}$ after 10 avalanche times $t\sub{ava}$, with an initial runaway-electron density $n\sub{RE,0} = 10^{10}\,$m$^{-3}$, $n_e = 5\times 10^{19}\,$m$^{-3}$ and $T = 100\,$eV. Runaway electrons in red and positrons in black and blue. Dashed lines denote the theoretical predictions of Eqs.~(\ref{eq:ss dist}) and (\ref{eq:rp dist}); in the (ii) $E = 10E_c$, $\Ztot = 1$ case it fully overlaps with the numerical solution.  \label{fig:dist}}
\end{figure}
Figure~\ref{fig:dist} illustrates the angle-averaged positron distribution for two cases: (\emph{i}) with $E=2E_c$ and $Z\sub{tot} = 10$, and (\emph{ii}) $E=10E_c$ and $Z\sub{tot}=1$.
%different plasma charges and electric-field strengths. 
The analytic solution, Eqs.~(\ref{eq:ss dist}) and (\ref{eq:rp dist}), is nearly indistinguishable from the numerical solution for $p<0$ for both cases, and for $p>0$ in case ($ii$) with the higher electric field (shown in blue). The analytic solution fails to fully capture the low energy-behaviour in case ($i$) with low electric field and high plasma charge (black), where pitch-angle dynamics become important. The accuracy of the analytic solution at high electric field further motivates the neglect of annihilation in the dynamics of fast positrons. The sharp peaks at $p=0$ in the numerical positron energy spectra contain the thermalized positron populations, which we do not consider the detailed dynamics of here.

\section{Rate equations for runaway positrons}
\label{sec:rate}
From the kinetic description of Section \ref{sec:source} we can find a reduced set of fluid equations which govern the evolution of the number densities of runaway positrons as well as thermal positrons.
We introduce the runaway positron density $n\sub{RP}$ and
%as the number
%density of positrons with positive $\xi$ (being born near $\xi \approx
%-1$ corresponding to the direction of runaway electrons) and momentum
%above the runaway threshold, $p_c > m_e c/\sqrt{E/E_c - 1}$, and the
thermal positron density $n\sub{TP}$ %as the number density of positrons with $p<p_c$:
in the same way as in the previous section.
These then satisfy the equations %%Geri: Your \Gammas are not really defined.
\begin{align}
\frac{\partial n\sub{RP}}{\partial t} &= Z\sub{tot} n_e n\sub{RE} \kappa(E,\,\Ztot ) \langle v\sigma_c^+\rangle\sub{RE} - n\sub{RP}/\tau\sub{aR}
 \label{eq:runaway positron fluid}  \\
%&= n\sub{RE} \Gamma^+- n\sub{RP}/\tau\sub{aR}, 
%\\
\frac{\partial n\sub{TP}}{\partial t} &= Z\sub{tot}n_e n\sub{RE}\eta(E,\,\Ztot ) \langle v\sigma_c^+\rangle\sub{RE} - n\sub{TP}/\tau\sub{aT}  
\label{eq:thermal positron fluid}\\
\frac{\partial n\sub{RE}}{\partial t} &= n\sub{RE}\Gamma\sub{ava}(E,\,\Ztot ).
\label{eq:runaway electron fluid}
\end{align}
where $\kappa$ denotes the fraction of created positrons that are
accelerated as runaways, $\eta$ the fraction that is thermalized, %$\Gamma^+$ and $\Gamma\sub{T}^+$ are the growth rate of the runaway positrons and thermal positrons, respectively,
and $\Gamma\sub{ava} = 1/ t\sub{ava}$ is the avalanche growth rate of
runaway electrons. 

\begin{figure}
\begin{center}
\includegraphics[width=0.49\textwidth]{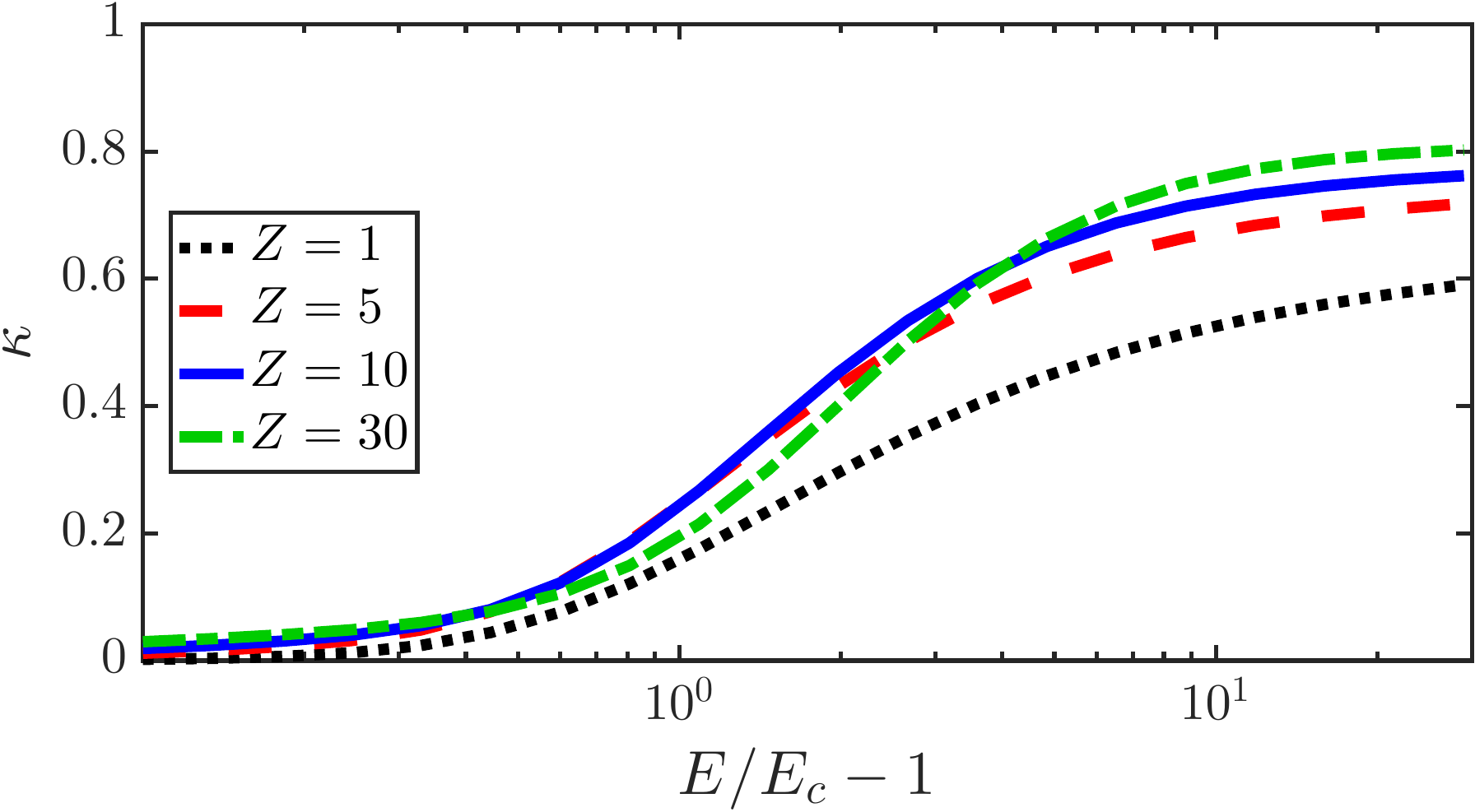}\hfill
\includegraphics[width=0.49\textwidth]{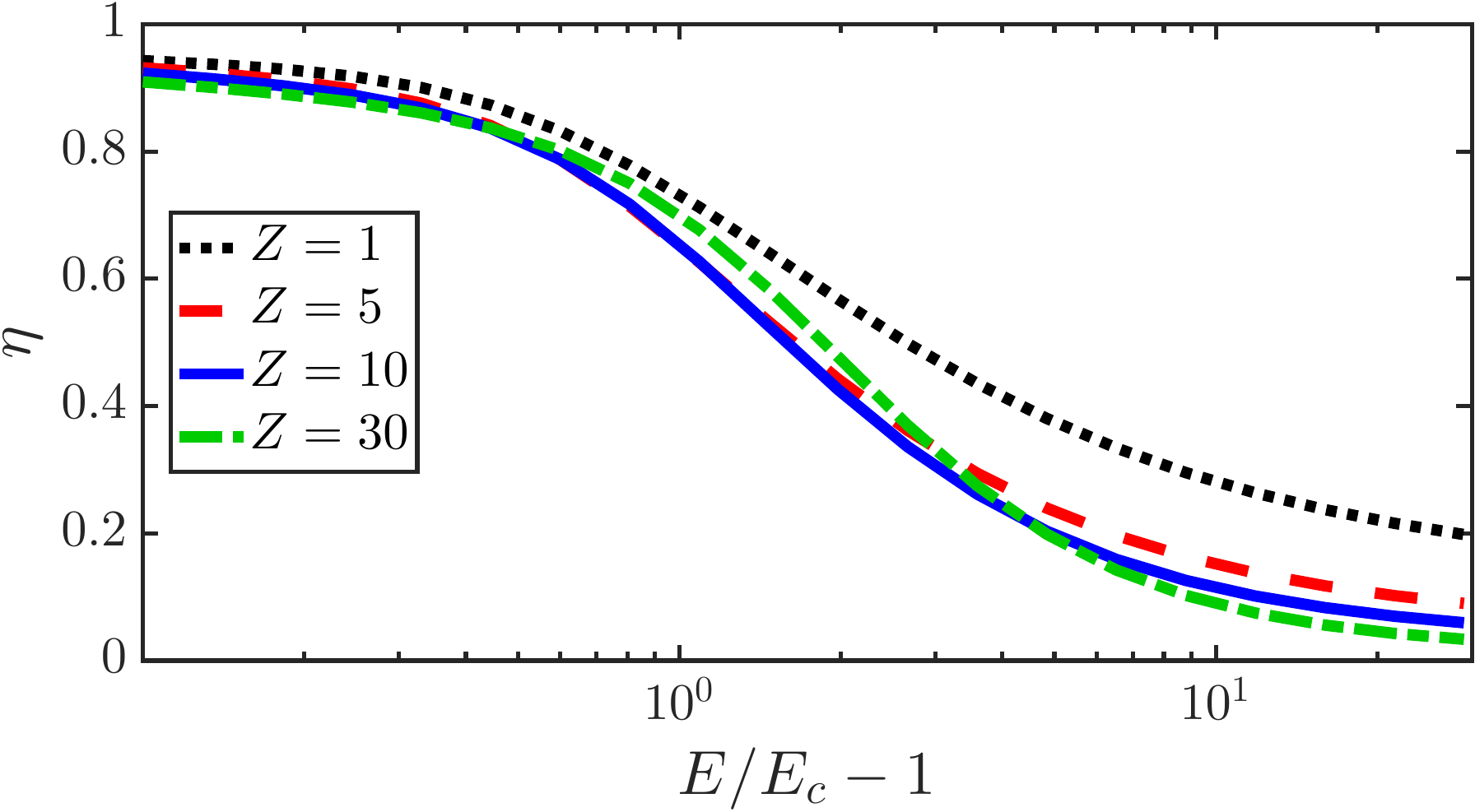}
\end{center}
\caption{\label{fig:runaway_fraction1}(left) Positron runaway fraction $\kappa$ defined by Eq.~(\ref{eq:runaway positron fluid}), for various electric-field strengths and plasma effective charge $\Ztot $. (right) Positron thermalization fraction $\eta$ defined by Eq.~(\ref{eq:thermal positron fluid}), for various electric-field strengths and plasma effective charge $\Ztot $. }
\end{figure} 

%\begin{figure}
%\begin{center}\includegraphics[width=0.6\textwidth]{eta_nRE_t_Ann}\end{center}
%\caption{\label{fig:runaway_fraction2}Positron thermalization fraction $\eta$ defined by Eq.~(\ref{eq:thermal positron fluid}), for various electric-field strengths and plasma effective charge $\Ztot $.}
%\end{figure}

Kinetic simulations must be used to determine the runaway fraction
$\kappa$ and thermalization fraction $\eta$ (note that they will not sum to unity, since the population of fast newly-born positrons with $\xi<0$ also grows in time). Results from numerical simulations for a variety of electric
fields and plasma charges are shown in Fig.~\ref{fig:runaway_fraction1}, obtained for constant electric fields and plasma charge. These are applicable to scenarios where $E$ and $\Ztot $ vary slowly in time compared to the avalanche time. 
When $E\gg
E_c$, the runaway fraction is near unity, but decreases exponentially
in magnitude when the electric field approaches the threshold value
$E_c$.  As only a small fraction of positrons are annihilated before
slowing down (or entering the runaway region) \citep{heitler}, the
effect of annihilation on the positron-runaway-generation dynamics can
be ignored, and we can assume that positrons are only annihilated
after being either thermalized or runaway-accelerated.

In the presence of a constant electric field, background density and charge, the rate equations have a simple analytic solution given by (after a short transient phase on the scale of $t\sub{ava}$)
\begin{align}
n\sub{RE}(t) &= n_0 \exp(\Gamma\sub{ava} t), \nonumber \\
n\sub{RP}(t) &= n\sub{RE}(t)\frac{Z\sub{tot} \kappa n_e  \langle v\sigma_c^+\rangle}{\Gamma\sub{ava} + \tau\sub{aR}^{-1}}, \label{eq:positron eq densities} \\
n\sub{TP}(t) &= n\sub{RE}(t)\frac{Z\sub{tot} \eta n_e  \langle v\sigma_c^+\rangle}{\Gamma\sub{ava}+\tau\sub{aT}^{-1}}.\nonumber
\end{align} 
\blue{Note that the positron populations grow in time despite annihilation; this occurs due to the ever-increasing amplitude of the positron source, since the runaway electrons are avalanching.}

When the electric field is significantly above threshold one finds that $\Gamma\sub{ava} \gg \tau\sub{aR}^{-1}$ \blue{meaning that annihilation is negligible}, so that
\begin{align}
\frac{n\sub{RP}}{n\sub{RE}} \approx \frac{\kappa(E,\,Z\sub{tot})}{E/E_c-1} \frac{Z\sub{tot}\langle v\sigma_c^+\rangle}{4\pi c r_0^2/c_Z} 
 \approx \alpha^2c_Z Z\sub{tot} \frac{\kappa(E,\,Z\sub{tot})}{E/E_c - 1} \frac{\gamma_0-6.7}{60\pi}\blue{,}
%&\approx \frac{\kappa(E,\,Z)}{E/E_c - 1} \frac{7}{27\pi^2} \frac{\Ztot \alpha^2}{\ln\Lambda } \int_1^\infty \rd \gamma \,e^{-\gamma/c_Z\ln\Lambda} \ln^3\gamma  \nonumber \\
%&\sim \frac{\kappa(E,\,Z)}{E/E_c - 1} \frac{7\alpha^2}{27\pi^2}\Ztot  c_Z \ln(\gamma_0/\gamma_e) \left[ \ln^2(\gamma_0/\gamma_e) + \pi^2/2 \right],
\label{eq:runaway ratio}
\end{align}
where again $\gamma_0 = c_Z\ln\Lambda$. The electric-field
dependence is fully captured in the factor $\kappa/(E/E_c-1)$, which
takes its maximal value $\approx 0.2$ near $E\approx 2E_c$, only
weakly dependent on the charge $\Ztot $. With $\ln\Lambda = 15$,
we then find that the maximal ratio of runaway positrons to electrons
is $n\sub{RP}/n\sub{RE} \lesssim 8.5\cdot 10^{-7} Z\sub{tot}
c_Z(c_Z-0.45)$, \blue{which for a low-$Z$ plasma with $\Ztot = 1$ is approximately 
$4\times 10^{-6}$, and for a 
high-$Z$ plasma with 
$\Ztot =20$ of the order of $4\times 10^{-4}$}.  This means that the
runaway-positron synchrotron and hard X-ray (HXR) emission may be
challenging to distinguish from the radiation emitted by the runaway
electrons in a tokamak, since even a small fraction of reflected \blue{or scattered}
radiation from electrons or noise from other sources could drown out
the positron signal.

\section{Radiation from positrons in tokamak plasmas}
\label{sec:rad}
In the previous section we found that runaway positrons are less numerous than the runaway electrons by a factor smaller than approximately $10^{-4}$. This causes a direct measurement of runaway positrons in a laboratory plasma to be challenging, and an appealing option is  instead to detect the annihilation radiation of the positrons that have slowed down, which is distinctly peaked around photon energies of 511\,keV. The annihilation radiation from slow positrons is emitted approximately isotropically, whereas runaway electrons emit radiation primarily along their direction of motion, which when the electric field is large is along the electric field, or along the magnetic field in a magnetized plasma. This means that when measuring perpendicularly to the direction of runaway acceleration, even though the positrons are much fewer, their annihilation radiation may be detected through the X-ray background of runaway electrons for which only a small fraction is emitted at a $\pi/2$ angle and near $511\,$keV\footnote{More precisely, the ratio of perpendicular to tangential bremsstrahlung emission is given by approximately $3/(8\gamma^4)$ at $k=511\,$keV.}. Furthermore, coincidence measurement techniques can be employed to carry out measurements in poor signal-to-noise ratio cases \citep{guanying}.

We can make the heuristic discussion above stricter by the following arguments. The number density of bremsstrahlung photons emitted per unit solid angle, time and photon energy is given by
\begin{align}
\frac{\partial n\sub{HXR}}{\partial t \partial \Omega \partial k} &= n_e \Ztot  \int_{\gamma > k+1} v \frac{\partial \sigma\sub{br}}{\partial k \partial \Omega} f\sub{RE}(\g{p}) \,\rd \g{p} .
\end{align}
This can be compared to the number density of annihilation photons emitted per unit time and solid angle due to the thermal positrons $n\sub{TP}$ annihilating against the cold background,
\begin{align}
&\frac{\partial n\sub{an}}{\partial t \partial \Omega} = \frac{n\sub{TP}}{4\pi\tau\sub{aT}} \approx Z\sub{tot}\frac{n\sub{RE}n_e\eta\langle v\sigma_c^+\rangle\sub{RE}}{4\pi} ,  
%&\approx \frac{7\alpha^2n\sub{RE}}{27\pi^2}\eta  n_e r_0^2 c \Ztot  \ln\frac{\gamma_0}{\gamma_e} \left[\ln^2\frac{\gamma_0}{\gamma_e} + \frac{\pi^2}{2}\right],
\end{align}
where we have assumed the thermal positron-annihilation rate to be much larger than the avalanche growth rate, $\Gamma\sub{ava}\tau\sub{aT} \ll 1$.

The annihilation radiation will have a line profile in photon energy with a characteristic width comparable to the background temperature. We consider the case where the profile is not resolved in the measurement, and the full line is captured in one channel. In this case, since the hard X-rays have a broad spectrum, we find it useful to characterize the visibility of the annihilation line with the parameter
\begin{align}
\Delta k = \frac{ {\partial n\sub{an}}/{\partial t \partial \Omega } }{ {\partial n\sub{HXR}}/{\partial t\partial \Omega \partial k} },
\end{align}
which (when $\Delta k \ll k$) can be interpreted as the photon-energy interval $\Delta k$ around $k = m_e c^2$ within which the total HXR emission equals the annihilation photon flux. From a detection point of view, $\Delta k$ would approximate the energy resolution required for the annihilation peak to appear with twice the amplitude of the continuous X-ray background. \blue{The finite line width of the annihilation peak would need to be accounted for when the plasma temperature satisfies $T \gtrsim \Delta k$.}

\begin{figure}
\begin{center}
\includegraphics[width=0.6\textwidth]{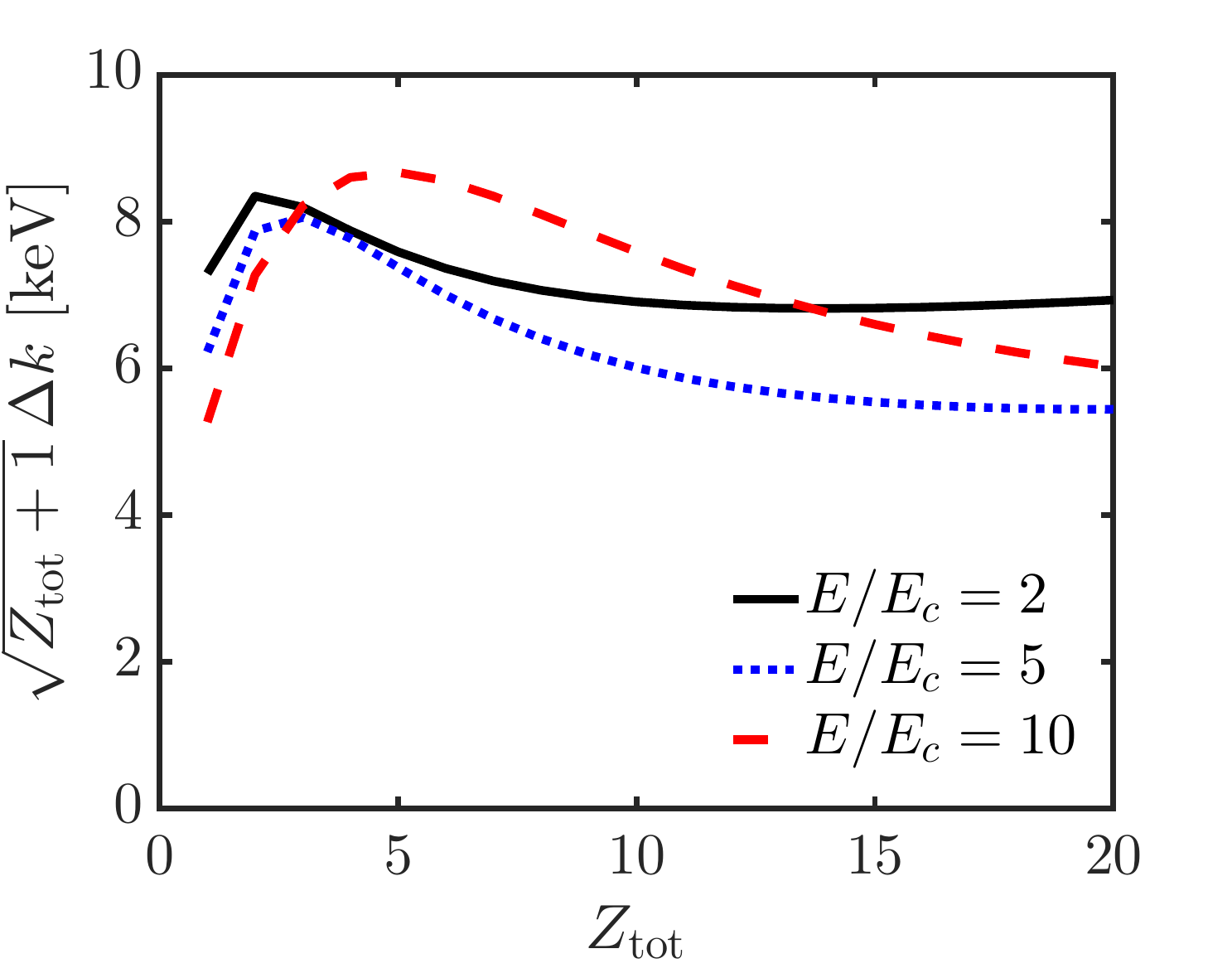}
\caption{\label{fig:delta k} The photon-energy resolution parameter $\Delta k = (\partial n\sub{an}/\partial t\partial \Omega)/(\partial n\sub{HXR}/\partial t \partial \Omega \partial k)$ for perpendicular detection of annihilation radiation from thermalized positrons and hard X-rays from runaway electrons.   }
\end{center}
\end{figure}

In Fig.~\ref{fig:delta k} we show $\Delta k$ for detection at a $\pi/2$ angle relative to the direction of runaway acceleration, using the analytic runaway distribution of Eq.~(\ref{eq:full RE dist}). We observe a relatively weak dependence on electric field where the main trend is approximately captured, within roughly 25\%, by
\begin{align}
\Delta k \approx \frac{7\,\text{keV}}{\sqrt{\Ztot+1} }.
\label{eq:delta k ROT}
\end{align}
This means that in order for the annihilation peak to be clearly distinguishable from the X-ray background due to runaway electrons, an energy resolution better than or comparable to $(7/\sqrt{\Ztot+1} )\,$keV is desirable.
The contrast of X-rays to annihilation radiation, quantified through $\Delta k$, is largely insensitive to other parameters of the scenario, since the cross sections for the two processes scale in the same way with the background parameters.

There are two main competing effects which are sensitive to $E$ and $\Ztot $ that determine the observed behavior in $\Delta k$. When $E$ increases, the thermalization fraction $\eta$ of positrons rapidly decreases, as illustrated in Fig.~\ref{fig:runaway_fraction1} (right), which reduces the amount of 511~keV annihilation radiation. At the same time the runaway-electron population becomes more anisotropic, which sharply reduces the amount of bremsstrahlung emitted at a perpendicular angle. In the parameter range shown in the figure, these effects are found to approximately cancel, leaving only a weak $E$ dependence. 
On the other hand, an increase in charge $\Ztot $ causes the electron population to become more isotropic, \emph{increasing} the amount of bremsstrahlung emitted at a perpendicular angle, however, it also increases the average runaway-electron energy which increases the number of positrons created per electron. The former effect is significantly stronger, which causes a net $1/\sqrt{\Ztot+1}$ dependence.

Finally we note that, in the post-disruption runaway plateau where the runaway current slowly dissipates on the inductive time-scale of the device, the analytical avalanche runaway distribution that we have used here is not valid, as it tends to significantly overestimate the average energy of the distribution. Due to its experimental accessibility, we  consider this scenario separately for a singly ionized argon-dominated plasma~\citep{ASDEX2016}. For the runaway electron distribution we use the self-consistent slowing-down distribution of Ref.~\citep{Hesslow2018} obtained from a numerical solution of the kinetic equation with an inductive electric field and accurate modelling of screening effects on collisions, at a plasma temperature $T=10$\,eV.
Using such a numerical distribution function in evaluating the rate of pair production $\langle v\sigma_c^+ \rangle\sub{RE}$ and the bremsstrahlung photon flux yields $\Delta k = 0.31$~keV, which is approximately $20\%$ of the value predicted by the rule-of-thumb given in Eq.~(\ref{eq:delta k ROT}) evaluated with $\Ztot  = 18$.

As well as being distinguishable from the runaway X-rays, it is required that the total number of annihilation photons reaching a detector is sufficiently large. While this is highly sensitive to the details of the setup, we can provide a rough estimate in the following way.
The total number of annihilation photons per unit time reaching a detector with area $A\sub{det}$
placed a distance $R$ from the plasma detecting emission within an opening angle $\Delta \theta$, is given approximately by 
\begin{align}
\frac{\partial N\sub{an}}{\partial t}& \approx  \Delta \theta  \frac{A}{R} A\sub{det} \frac{\partial n\sub{an}}{\partial t \partial \Omega} \nonumber \\
&\approx \Delta\theta  A\sub{det}A Z\sub{tot} \frac{n\sub{RE}n_e\eta\langle v\sigma_c^+\rangle\sub{RE}}{4\pi R}
\nonumber \\
&\approx \Delta\theta  A\sub{det}\frac{ A n\sub{RE} ec}{e} Z\sub{tot} \frac{n_e r_0^2 }{4\pi 137^2 R}\eta \frac{\gamma_0-6.7}{15}
\nonumber\\
& \approx (1.4\times  10^{6}\,\text{s}^{-1}) n_{20}I\sub{RE}[100\,\text{kA}]\times \nonumber \\
&\hspace{10mm}\times \Delta\theta\frac{A\sub{det}[\text{dm}^2]}{R[\text{m}]}Z\sub{tot} (\gamma_0-6.7)\eta.
%\nonumber\\
%&= ( 34.5 \,\text{ms}^{-1})  n_{20} I\sub{RE}[100\,\text{kA}] \frac{\Delta \theta}{0.5\,\text{rad}}\frac{A\sub{det}[\text{cm}^2]}{R[\text{m}]} \times \nonumber \\
%&\times  \Ztot \,\eta(\Ztot ,E/E_c)  \ln\frac{\gamma_0}{\gamma_e} \left[\ln^2\frac{\gamma_0}{\gamma_e} + \frac{\pi^2}{2}\right] \nonumber \\
%&\approx  ( 1\times 10^3 \,\text{[ms]}^{-1})  \Ztot  n_{20} I\sub{RE}[100\,\text{kA}] \frac{\Delta \theta}{0.5\,\text{rad}}\frac{A\sub{det}[\text{cm}^2]}{R[\text{m}]},
\end{align}
%where the last line is approximately valid when $E\approx E_c$ so that most created positrons slow down. 
Here, the cross-sectional area $A$ of the plasma is assumed to be completely within the detector field-of-view. Then, discharges with higher plasma charge, background density and runaway current are seen to yield higher total annihilation-photon fluxes. Note that a strong decrease in total photon flux is found when the electric field increases above the threshold value $E_c$, due to the change in the thermalization fraction $\eta$. 
As an example, inserting values typical of a disruption in a medium-sized tokamak with $R = 1.5\,$m, $Z\sub{tot} = 10$, $\ntot=10^{20}$\,m$^{-3}$, $I\sub{RE} = 400\,$kA, $E=2E_c$, $A\sub{det} = 1\,$dm$^2$, $\Delta \theta = 0.5\,$rad and with $\ln\Lambda = 15$, one obtains a detected 511\,keV annihilation photon count of $\partial N\sub{an}/\partial t \approx 7\times 10^8\,$s$^{-1}$.

In poor signal-to-noise ratio cases coincidence measurements can be employed, where only positrons annihilated between two detectors are counted. This can be approximately accounted for in the previous formula by using an opening angle $\Delta \theta = \sqrt{A\sub{det}}/R$ if two identical detectors are placed on either side of the plasma, which reduces the number of counts by another factor $0.1\,\sqrt{A\sub{det}}$[dm]/$R$[m].

\section{Conclusions}
\label{sec:concl}
Fast electrons can produce electron-positron pairs, primarily via either a
two-step process based on the emission of a bremsstrahlung photon and
a subsequent photon-particle interaction, or the direct process where
pairs are produced in collisions between fast electrons and
nuclei. We show that the former process is dominant when the electric
field is above a certain threshold value, which is given in equation (\ref{eq:threshold field}) 
and illustrated in Fig.~\ref{fig:critical pp field}. 
The latter process is however always dominant when the fast electrons are
confined to a region in space which is smaller than the photon
mean-free path, e.g.~in magnetic fusion plasmas.  
Using a differential cross section for collisional pair production 
calculated using MadGraph 5~\citep{madgraph}, it is revealed that previous
studies of pair production during runaway have significantly overestimated 
the positron generation rate.

In strong electric fields electrons and positrons are accelerated and
may run away. The kinetic equations for electrons and positrons are
similar, except for the opposite directions of acceleration in an electric field,
and the source and annihilation terms present in the positron
kinetic equation. We show that when the electric field is
sufficiently large the positron distribution function can be
calculated analytically, with explicit solutions given in Eqs.~(\ref{eq:ss
  dist}) and (\ref{eq:rp dist}). The analytical solution shows
remarkable agreement with numerical solutions of the kinetic
equation in the relevant limit (high electric field and moderate
charge number), as illustrated in Fig.~\ref{fig:dist}.

Since the characteristic initial energy of the newly born positrons is
large, a fluid description for the positron population can be
used. Kinetic simulations are then only needed to determine the
fraction of created positrons that are thermalized or
runaway-accelerated as a function of the background parameters. The
evolution of the number density of thermal and runaway positrons can
then be calculated from simple rate equations, given in
Eqs.~(\ref{eq:runaway positron fluid})-(\ref{eq:runaway electron
  fluid}).  These equations admit analytical solutions in the presence
of a constant electric field, and can be used to determine the ratio
of the runaway positron and electron populations.  The runaway and
thermalized positron fractions determined from numerical kinetic
simulations are given for a variety of electric fields and charge
numbers in Fig.~\ref{fig:runaway_fraction1}.

Finally we calculate the radiation emitted by a positron population in
a post-disruption tokamak plasma, and evaluate the annihilation to
HXR ratio of photon fluxes emitted at a perpendicular angle to the
system. Using these, one can estimate the parameter region where
positrons can be detected, that is when their annihilation radiation
is not overwhelmed by the bremsstrahlung radiation of energetic
electrons and when the total photon count is sufficiently large.

\acknowledgments The authors would like to thank G Ferretti and
I Pusztai for fruitful discussions.  This work was
supported by the European Research Council (ERC-2014-CoG grant 647121)
and the Swedish Research Council (Dnr.~2014-5510).

\appendix

\section{Positron source term}
\label{ap:source}
The differential cross sections appearing in Eq.~(\ref{eq:full cs}) are given in the Born approximation by
\citep{heitler}
\begin{align}
%\frac{\rd \sigma^+_{c}}{\partial \gamma} &= \frac{56 r_0^2
%\alpha^2}{9\pi} \frac{\ln \gamma}{\gamma} \big[\ln \gamma_1 -
%\ln \gamma \big], \label{eq:sigma_c}\\
\frac{\partial \sigma_\gamma^+}{\partial \gamma} &= \alpha r_0^2 \frac{p p_-}{k^3}\Biggr\{ -\frac{4}{3} - 2 \gamma_-\gamma\frac{p_-^2 + p^2}{p_-^2p^2}+ \frac{\gamma}{p_-^3}\epsilon_- \nonumber \\
& + \frac{\gamma_-}{p^3} \epsilon - \frac{\epsilon_-\epsilon}{p_-p} + L_-\biggr[ k^2\frac{\gamma_-^2\gamma^2 + p_-^2p^2}{p_-^3 p^3}-\frac{8}{3}\frac{\gamma_-\gamma}{p_-p} \nonumber \\
&- \frac{k}{2p_-p}\biggr(\frac{\gamma_-\gamma-p_-^2}{p_-^3}\epsilon_- + \frac{\gamma_-\gamma - p^2}{p^3}\epsilon + \frac{2k\gamma_-\gamma}{p_-^2p^2}\biggr)\biggr]\Biggr\}  ,\\
\frac{\partial \sigma\sub{br}}{\partial k} &= \alpha r_0^2 \frac{p}{kp_1} \Biggr\{
    \frac{4}{3} - 2\gamma_1\gamma\frac{p_1^2+p^2}{p_1^2p^2} + \epsilon_1\frac{\gamma}{p_1^3}+\epsilon\frac{\gamma_1}{p^3} - \frac{\epsilon_1\epsilon}{p_1p} \nonumber \\
    &+L_1\Biggr[ \frac{8}{3}\frac{\gamma_1\gamma}{p_1p}  +k^2\frac{\gamma_1^2\gamma^2+p_1^2p^2}{p_1^3p^3} +\frac{k}{2p_1p}\biggr( \epsilon_1\frac{\gamma_1\gamma+p_1^2}{p_1^3}  \nonumber \\
    &- \epsilon\frac{\gamma_1\gamma+p^2}{p^3} + 2k\frac{\gamma_1\gamma}{p_1^2p^2}  \biggr) \Biggr] 
    \Biggr\}, \nonumber\\\
     \epsilon &= 2\ln(\gamma+p) \\e
     \epsilon_1 &= 2\ln(\gamma_1+p_1) \nonumber \\
     \epsilon_- &= 2\ln(\gamma_- + p_-) \nonumber \\
     L_1 &= 2\ln\frac{\gamma_1\gamma+p_1p-1}{k}, \nonumber  \\
     L_- &=2\ln\frac{\gamma_-\gamma + p_-p-1}{k}, \nonumber 
\end{align}
and $1/\alpha = 4\pi\varepsilon_0 \hbar
c/e^2 \approx 137$ denotes the inverse fine-structure constant. In the expression for $\partial \sigma^+_\gamma/\partial \gamma$, energy conservation constrains $\gamma_- = k-\gamma$ where $p_- = \sqrt{\gamma_-^2-1}$ denotes the momentum of the electron created in the pair, whereas in the expression for $\partial \sigma\sub{br}/\partial k$ one has $\gamma = \gamma_1-k$ and $p = \sqrt{\gamma^2-1}$ is the momentum of the outgoing positron. Here, momenta are expressed in units of $m_e c$ and $k$ is the photon energy in units of $m_e c^2$.

\begin{figure}
  \begin{center}
\includegraphics[width=0.6\textwidth,trim=0mm 30mm 0mm 0mm]{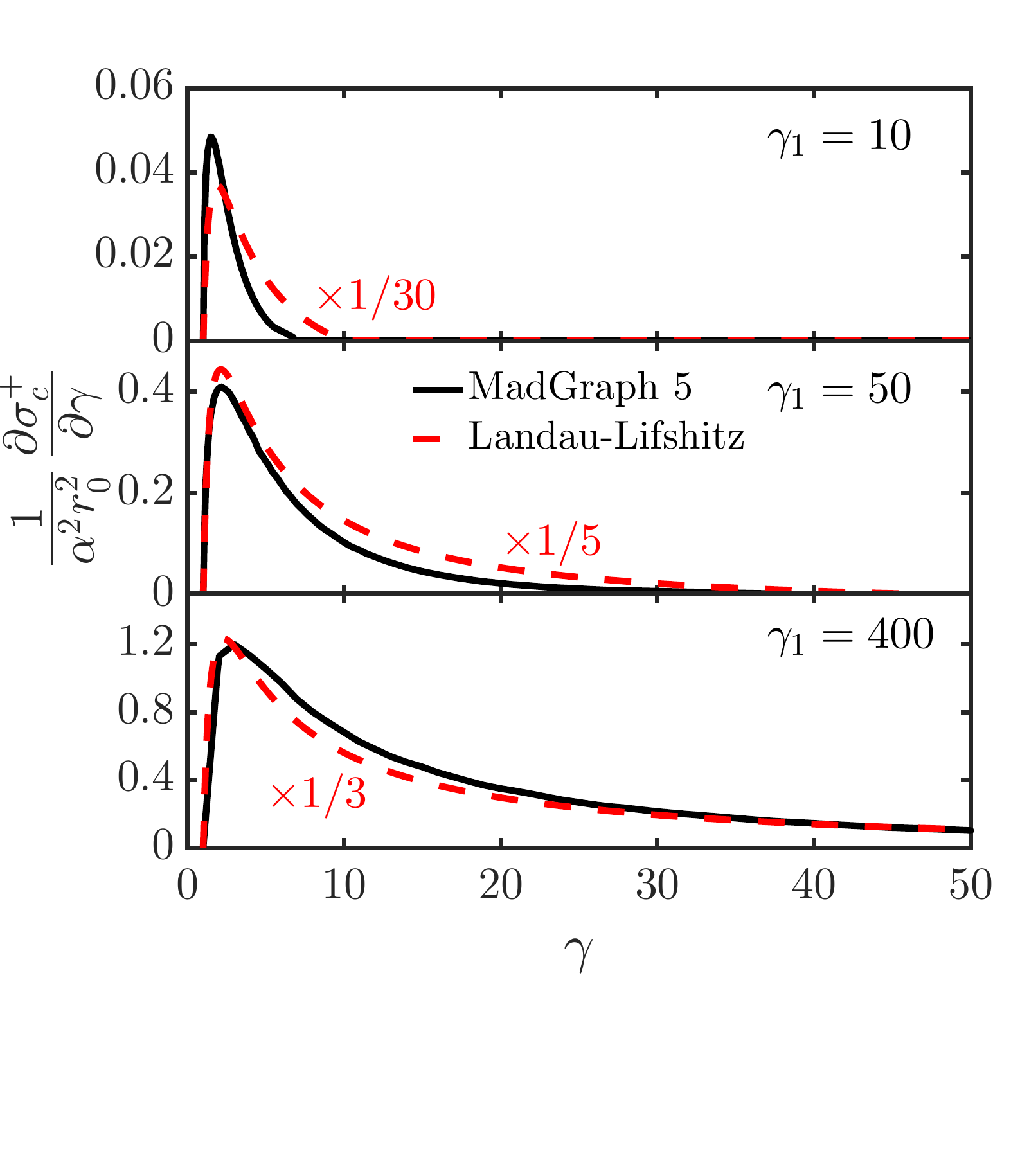}\end{center}
\caption{\label{fig:diffCrossSecComp} Differential cross section for pair production in collisions, by MadGraph~5 (solid black line, employed for results in this paper) and for comparison the Landau-Lifshitz formula (dashed red). \blue{$\gamma_1$ and $\gamma$ are the incident-electron and outgoing-positron Lorentz factors, respectively.} The Landau-Lifshitz formula has been multiplied by 1/30, 1/5 and 1/3 in the three subplots, respectively, in order to illustrate better the shapes of the curves. The MadGraph~5 results are significantly overestimated by the approximate formula.}
\end{figure}

The cross section $\partial \sigma_c^+/\partial \gamma$ for pair production in collisions by electrons and ions is evaluated in the Born approximation by the MadGraph~5 tool~\citep{madgraph}, where 1,300,000 events were generated for each incident electron energy $\gamma_1$, for which 140 values between 3.13 and 587 were sampled (corresponding to a range from 1.6\,MeV to 300\,MeV). In Fig.~\ref{fig:diffCrossSecComp} we compare $\partial \sigma_c^+/\partial \gamma$ as calculated by MadGraph~5 with the corresponding differential cross section evaluated in the main logarithmic approximation neglecting contributions of order $1/\ln\gamma_1$~\citep{landauQED},
\begin{align}
\frac{\partial \sigma^+_{c,\text{LL}}}{\partial \gamma} = \frac{56 \alpha^2 r_0^2}{9\pi}\frac{\ln\gamma}{\gamma}\ln\frac{\gamma_1}{\gamma}.
\end{align}
We observe that the shape of the Landau-Lifshitz cross section $\rd \sigma^+_{c,\text{LL}}$ is qualitatively similar to the MadGraph~5 results, although the values deviate significantly from the more accurate calculation. At moderate-to-low electron energies, the Landau-Lifshitz formula also significantly overestimates the average positron energy. \blue{The disagreement between the Landau-Lifshitz formula and the corresponding Born approximation result is expected, since the logarithmic approximation is only valid at significantly higher energies than those relevant to runaway scenarios.}

\begin{figure}
  \begin{center}
\includegraphics[width=0.6\textwidth,trim=0mm 0 0mm 0]{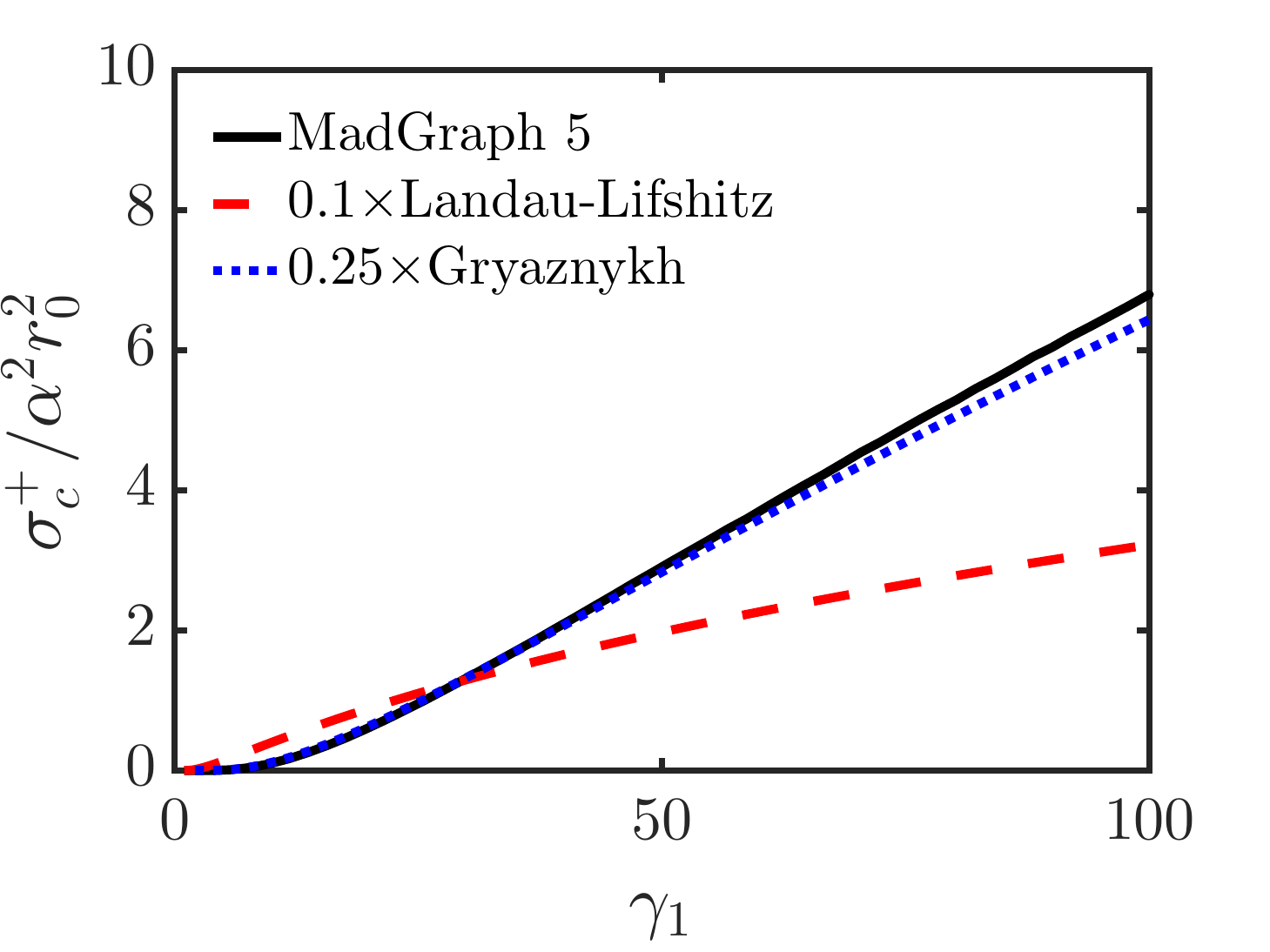}\end{center}
\caption{\label{fig:totalCrossSecComp} Total cross sections $\sigma_c^+$ by MadGraph~5 (solid, black), the Landau-Lifshitz formula (dashed, red) and the Gryaznykh formula (dotted, blue) \blue{as function of the incident-electron Lorentz factor $\gamma_1$}. The Landau-Lifshitz and Gryaznykh formulas have been rescaled for better visibility; they both significantly overestimate the positron production compared to the full MadGraph~5 computation.}
\end{figure}

In Fig.~\ref{fig:totalCrossSecComp} we compare the total pair production cross section $\sigma_c^+$ between MadGraph~5, the Landau-Lifshitz formula as well as with the formula given by~\cite{gryaznykh1998},
\begin{align}
\sigma^+_{c,\text{Gr}} = (5.11\,\mu\text{b}) \ln^3\frac{\gamma_1+3.6}{6.6},
\end{align} 
that has been employed in previous runaway positron studies. Although Gryaznykh's formula is a numerical fit to the full Born approximation result, it appears that the prefactor is too large by a factor of 4.

\begin{figure}
  \begin{center}
\includegraphics[width=0.6\textwidth,trim=5mm 0 5mm 0]{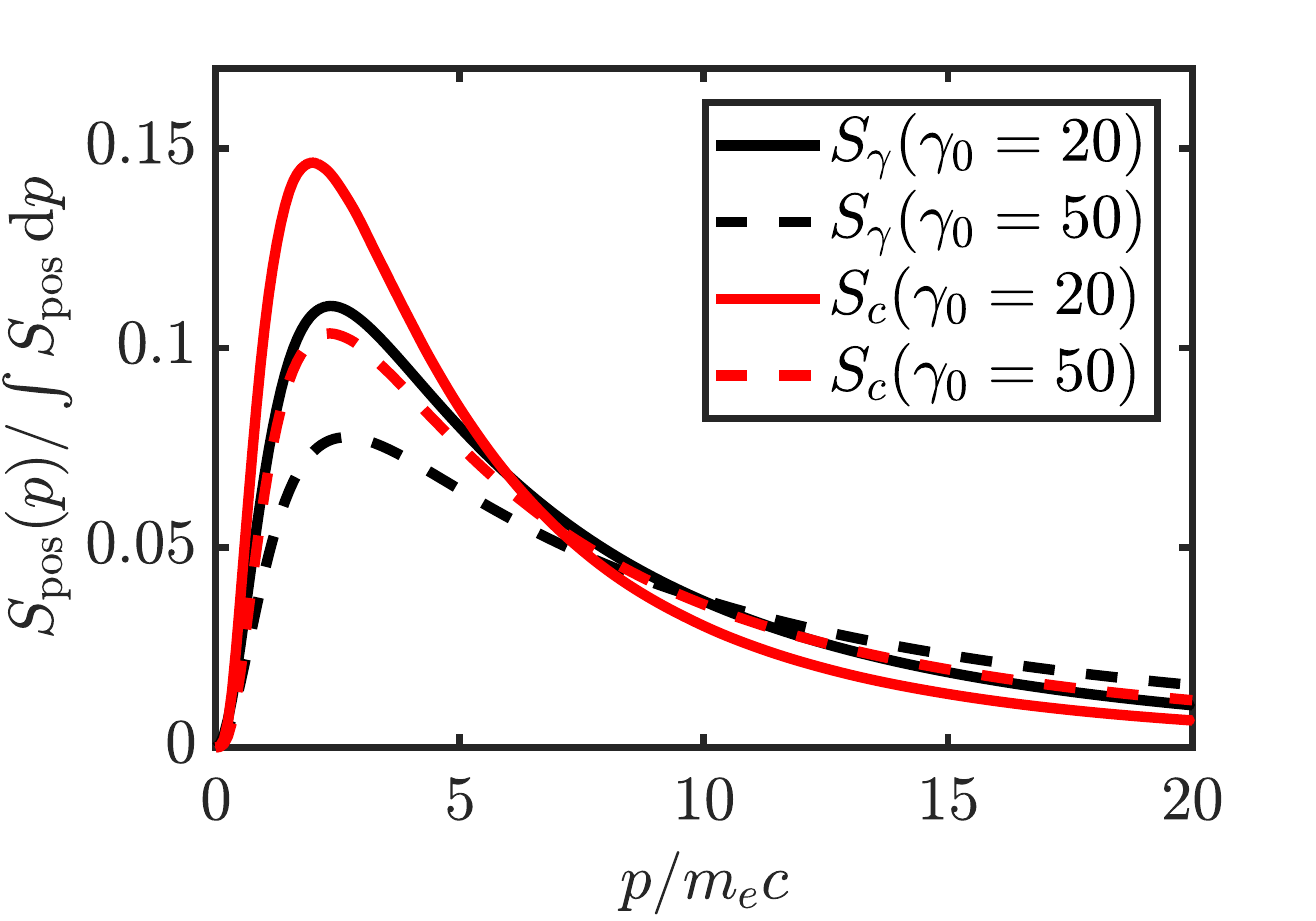}\end{center}
\caption{\label{fig:angle_averaged_source} Positron source terms $S_c$ (red), due to collisional pair production, and $S_\gamma$ (black), due to pair production via X-ray emission, normalized to unity production rate, evaluated at different values of \blue{the average runaway energy} $\gamma_0 = c_Z\ln\Lambda$. (solid, $\gamma_0 = 20$; dashed $\gamma_0 = 50$)}
\end{figure}

It is furthermore insightful to consider the energy spectrum of created positrons
by integrating the positron source $S\sub{pos}$ of Eq.~(\ref{eq:full source}) over angles,
\begin{align}
S\sub{pos}(p) = p^2\int  S\sub{pos}({\g p}) \, \rd \Omega_{\g p} = S_c(p) + S_\gamma(p),
\end{align}
where we have split the source into the contribution $S_c$ from collisional pair production and $S_\gamma$ from pair production via X-rays. These are defined so that $\int S\sub{pos}\,\rd p$ is the total rate at which positrons are produced, and are given explicitly by
\begin{align}
S_c &= n_e \Ztot  v \int_{\gamma+2}^\infty\rd\gamma_1 \,\frac{\partial \sigma_c^+}{\partial \gamma}\mathcal{F}\sub{RE}(\gamma_1), \\
S_\gamma &= n_e^2 \Ztot  t\sub{ava} v c \int_{\gamma+2}^\infty \rd \gamma_1 \int_{\gamma+1}^{\gamma_1-1}\rd k \, \frac{\partial \sigma_\gamma^+}{\partial \gamma}\frac{\partial \sigma\sub{br}}{\partial k} \mathcal{F}\sub{RE}(\gamma_1).
\end{align}
%Using the electron distribution (\ref{eq:1D RE dist}),
%one can analytically evaluate the collision term $S_c$ in the limit $\gamma \gg 1$,
%\begin{align}
%S_c(p) &\approx \frac{56 r_0^2 \alpha^2n_e \Ztot c}{9\pi }
%\frac{n\sub{RE}}{c_z \ln{\Lambda} }\frac{\ln\gamma}{\gamma^2
%  p}\times\int_p^\infty \rd p_1 \,
%\ln{\big(\frac{\gamma_1}{\gamma}\big)}\exp\left (-\frac{\gamma_1}{c_z
 % \ln{\Lambda}} \right ) \nonumber\\ &= \frac{56 r_0^2 \alpha^2n_e
 % n\sub{RE} \Ztot c}{9\pi}\frac{\ln\gamma}{\gamma^2
%  p}\Gamma\left(0, \frac{\gamma}{c_z \ln\Lambda}\right),
%\end{align}
%where $\Gamma(0,x) = \int_x^\infty e^{-s}/s \, \rd s$ is the
%incomplete gamma function. 
Figure~\ref{fig:angle_averaged_source} shows $S_c$ and $S_\gamma$ for two different systems, characterized by $\gamma_0\equiv c_Z\ln\Lambda = 20$ and $\gamma_0= 50$. It illustrates the dependence on the positron momentum $p$ of the two pair production mechanisms, when averaged over the electron (and photon) distribution. It is clear that the two main pair-production channels due to runaway electrons -- in collisions and via X-rays -- produce very similar positron energy spectra. 

We find that the average positron energy is not particularly sensitive to the average electron energy $\gamma_0$: by evaluating $\langle \gamma \rangle = \int_0^\infty \gamma S \,\rd p / \int_0^\infty S \,\rd p$, we obtain $\langle \gamma \rangle_c \approx 8$ and $11$ when $\gamma_0=20$ and 50, respectively, for the collision term $S_c$. For the X-ray term $S_\gamma$ we find $\langle \gamma \rangle_\gamma \approx 9$ and $13$ for the corresponding cases. Energies of newly created positrons during runaway are therefore typically always in the 5\,MeV range on average.

\section{Derivation of positron distribution function}
\label{ap:distributions}
We here present the derivation of the positron distributions (\ref{eq:ss dist}) and (\ref{eq:rp dist}) in the high-energy, small-pitch-angle limit.
The positron distribution varies over energies much larger than the rest energy, and thus satisfies the ultra-relativistic, one-dimensional kinetic equation
\begin{align}
\frac{\partial \mathcal{F}(p,\,t)}{\partial t} + eE_c\left(\frac{E}{E_c}-\text{sgn}(p)\right)\frac{\partial \mathcal{F}(p,\,t)}{\partial p}
= n_e c\Ztot \int_{\gamma+2}^{\infty} \rd \gamma_1 \, {\frac{\partial \sigma}{\partial \gamma}\hspace{-0.8mm}}^+\hspace{-1.5mm}(\gamma,\,\gamma_1) \mathcal{F}\sub{RE}(p_1,\,t),
\end{align}
where $p_1 = \text{sgn}(p)\sqrt{\gamma_1^2-1}$.
Here we have ignored the effect of annihilation on the evolution of the distribution, since this process occurs on a significantly longer time scale than the acceleration time in a strong electric field, and neglected the weak logarithmic energy dependence in the collisional friction force, taken to be constantly of magnitude $eE_c = m_e c/\tau_c$ with $\tau_c^{-1} = 4\pi \ln\Lambda n_e r_0^2 c$, opposing the direction of motion. \blue{We furthermore neglect radiation losses through synchrotron emission and bremsstrahlung; for high $Z$ and low $E$, this assumption may be violated in the far tail of the energy distribution.}
%We also consider only light elements for which bremsstrahlung radiation losses are negligible for energies in the tens of MeV range, and assume that any magnetic field present is sufficiently weak that the energy loss through synchrotron emission can be ignored.

Since the electron population\,--\,which drives the generation of positrons through the pair-production source term\,--\,grows exponentially in time we expect 
the positron kinetic equation to have a quasi-steady-state solution of the form $\mathcal{F}(p,\,t) = e^{t/t\sub{ava}}\mathcal{F}(p,\,0)$, growing at the same rate as the energetic electrons. For $p<0$, the positron distribution then satisfies the first-order linear ODE 
\begin{align}
\biggr[\frac{E/E_c-1}{\gamma_0} - \left(\frac{E}{E_c}+1\right) \frac{\partial}{\partial \gamma}\biggr]\mathcal{F} 
= \frac{\Ztot }{4\pi\ln\Lambda r_0^2}\frac{n\sub{RE}}{\gamma_0 m_e c}\int_\gamma^\infty \rd \gamma_1 \, \frac{\partial \sigma^+}{\partial \gamma} e^{-\gamma_1/\gamma_0},
\end{align}
where $\gamma_0 = c_Z\ln\Lambda$ is the average runaway-electron energy.
Imposing the boundary condition $\mathcal{F}(-\infty,\,t) = 0$, thus constraining the solutions to a finite total positron number, it is solved by
\begin{align}
\mathcal{F} =  \frac{\Ztot }{4\pi\ln\Lambda r_0^2}\frac{n\sub{RE}}{\gamma_0 m_e c} e^{\rho\gamma/\gamma_0}  
 \times \int_\gamma^\infty \rd \gamma' \int_{\gamma'+2}^\infty \rd \gamma_1 \, \Biggr[\frac{\partial \sigma^+}{\partial \gamma'}(\gamma',\gamma_1)\exp\left\{-\frac{\rho\gamma'+\gamma_1}{\gamma_0}\right\}\Biggr]
\end{align}
where $\rho = ({E/E_c-1})/({E/E_c+1})$.

Conversely, for $p>0$, the pair-production source vanishes, and the positron distribution satisfies the same equation as the high-energy runaway electrons except for the opposite charge,
\begin{align}
\frac{\partial \mathcal{F}}{\partial t} + eE_c\left(\frac{E}{E_c}-1\right)\frac{\partial \mathcal{F}}{\partial p}  = 0,
\end{align}
which is solved by 
\begin{align}
\mathcal{F}(p,t) = \mathcal{F}\left(0,\,t - \frac{p}{e(E -E_c)}\right).
\end{align}
Using as boundary condition at $p=0$ that the positron population grows in time in the same way as the $p<0$ population:
\begin{align}
\frac{\mathcal{F}(0^+,\,t)}{\mathcal{F}(0^+,\,0)} = \frac{\mathcal{F}(0^-,\,t)}{\mathcal{F}(0^-,\,0)} = e^{t/t\sub{ava}},
\end{align}
which then immediately yields
\begin{align}
\mathcal{F}(p,t) &= \frac{n\sub{RP}(0)}{m_e c \gamma_0} e^{t/t\sub{ava}}e^{-\gamma/\gamma_0}.
\end{align}

\bibliographystyle{jpp}
\bibliography{references}
\end{document}